\documentstyle[12pt,twoside,psfig]{article}
\begin{document}
\begin{center}{\Large\bf Study of non-canonical scalar field model using various parametrizations of dark energy equation of state }
\\[15mm]
Abdulla Al Mamon\footnote{E-mail : abdullaalmamon.rs@visva-bharati.ac.in}~and
Sudipta Das\footnote{E-mail:  sudipta.das@visva-bharati.ac.in}\\
{\em Department of Physics, Visva-Bharati,\\
Santiniketan- 731235, ~India.}\\
[15mm]
\end{center}
\vspace{0.5cm}
\vspace{0.5cm}
\pagestyle{myheadings}
\newcommand{\be}{\begin{equation}}
\newcommand{\ee}{\end{equation}}
\newcommand{\bea}{\begin{eqnarray}}
\newcommand{\eea}{\end{eqnarray}}
\newcommand{\bc}{\begin{center}}
\newcommand{\ec}{\end{center}}
\begin{abstract}
In this present work, we try to build up a cosmological model using a non-canonical scalar field within the framework of a spatially flat FRW space-time. In this context, we have considered four different parametrizations of the equation of state parameter of the non-canonical scalar field. Under this scenario, analytical solutions for various cosmological parameters have been found out. It has been found that the deceleration parameter shows a smooth transition from a positive value to some negative value which indicates that the universe was undergoing an early deceleration followed by late time acceleration which is essential for the structure formation of the universe. With these four parametrizations, the future evolution of the models are also discussed. It has been found that one of the models (Generalized Chaplygin gas model, GCG) mimics the concordance $\Lambda$CDM in the near future, whereas two other models (CPL and JBP) diverge due to future singularity. Finally, we have studied these theoretical models with the latest datasets from SN Ia $+$ $H(z)$ $+$ BAO/CMB.
\end{abstract}
{\bf keywords:} Dark energy, Non-canonical scalar, Parametrization, Data analysis
\section{Introduction}
Recent cosmological observations \cite{1acc,2acc,3acc,4acc,5acc,6acc} strongly suggest that our universe is presently accelerating. In literature, there has been a number of theoretical models to explain the origin of this acceleration mechanism. In this context, the most accepted idea is that an exotic component of the matter sector with large negative pressure, dubbed as ``dark energy" (DE), is responsible for this accelerated expansion of the universe. DE also makes up about $73\%$ of the total energy budget of the universe at present epoch. However, understanding the origin and nature of DE is still a challenging problem in modern cosmology. A number of models have been proposed phenomenologically as DE models, such as quintessence (canonical scalar field) \cite{quint1,quint2,quint3}, phantom \cite{phantom1,phantom2,phantom3}, k-essence \cite{kess22,kess23,kess24}, Chaplygin gas \cite{cgas1,cgas2,cgas3}, $f(R)$-gravity models \cite{fr1,fr2,fr3,fr4,fr5} and so on. The simplest theoretical candidate of DE is the vacuum energy with a constant equation of state (EoS) parameter $\omega = -1$, but it suffers from cosmological constant problem \cite{ccp1,ccp2}. The dynamical nature of dark energy also introduces a new cosmological problem, namely, ``coincidence" problem \cite{cp}. One alternative to the coincidence problem are coupled dark energy models where DE interchanges energy with the dark matter (DM) by means of a coupling term \cite{interaction1,interaction2,interaction3,interaction4,interaction5,interaction6}. Though a number of theoretical models have been proposed, none of them provides a satisfactory solution to all the problems. Hence, there is still a need of an appropriate model to explain current observations.\\
\par In the proposed models of dynamical dark energy, the EoS parameter is usually considered to be evolving with time. In this context, a large number of parametrizations of DE equation of state have been proposed \cite{cpl0,cpl01,cpl1,cpl2,cpl3,ba,cpl5,cpl6,cpl7,cpl10,cpl20,jbp}, which could provide solutions to a number of cosmological problems. However, most of these analysis have been carried out for canonical scalar field models of DE. Recently, non-canonical scalar field models are also being studied as a candidate for DE. In the non-canonical scalar field models, the kinetic part of the scalar field is modified and it has been found that an accelerated expansion can be achieved through these modifications. Originally, Armendariz-Picon et al. \cite{ncm2,ncm2gv} proposed this scenario to explain inflation at high energies where the non-canonical scalar field efficiently plays the role of {\it inflaton} (for review on this topic, see \cite{ncm2,ncm2gv,ncm1,ncm1rj,ncm1db,ncm23,ncm231,ncm232,ncm233,ncm3,
ncm301,ncm302,unni,tgolan,Fang}). Later, Chiba et al. \cite{kess22} introduced this scenario for dark energy models. In our recent work \cite{nc}, we have studied an interacting non-canonical scalar field model with a constant EoS parameter for the scalar field. The model was however restricted in the sense that a constant EoS parameter does not provide a general framework. Motivated by the above facts, in this present work, we wish to consider a varying EoS parameter for the non-canonical field. We consider four popular DE parametrizations for a non-canonical scalar field model in order to explain the late-time scenario of the universe; the parametrizations considered are Chevallier-Polarski-Linder parametrization \cite{cpl10,cpl20},
Jassal-Bagla-Padmanabhan parametrization {\cite{jbp}}, Barboza-Alcaniz parametrization \cite{ba} and {Generalized Chaplygin Gas parametrization \cite{cgas1,cgas2,cgas3}.\\
\par The features of these various parametrizations have been discussed in details in the next section. We have obtained the expressions for different relevant cosmological parameters, such as the deceleration parameter, density parameters of the scalar field and matter field for each model, and have shown that it is possible to have late time accelerated expansion of the universe for each of these choices.  We have also compared the results with standard canonical scalar field models considering these parametrizations. Furthermore, in this paper, we have discussed about the future evolution of the universe for these EoS parametrizations and we have found that one of these parametrizations (GCG) behaves as standard $\Lambda$CDM in far future where as CPL and JBP models fail to provide information about the far future and are valid till  $z>-1$. BA model however can provide information regarding the entire evolution history of the universe till $z=-1$. Finally, we have studied the constraint on the EoS parameter using the combination of SN Ia + $H(z)$ + BAO/CMB dataset.\\
\par The paper is organized as follows. In section 2, we have described the basic theoretical framework for the non-canonical scalar field model of a flat FRW universe. We have then solved the governing dynamical equations for this toy model using four different types of DE parametrizations of the EoS parameter. It has been found that the resulting cosmological scenarios are in good agreement with the current observations in each case. In section 3, we have obtained the observational constraints on this model parameters using SN Ia + $H(z)$ + BAO/CMB dataset. Finally, some conclusions are presented in the last section. 
\section{Field equations and their solutions}
The general action for a scalar field model (with $8{\pi}G=c=1$) is given by
\be\label{action}
S = \int\sqrt{-g} d^{4}x\left[\frac{R}{2} + {\cal L}(\phi,X)\right] + S_{m}
\ee
where $R$ is the Ricci scalar curvature, ${\cal L}(\phi,X)$ is the Lagrangian density which is an arbitrary function of the scalar field $\phi$ and its kinetic term $X$. The kinetic term $X$ is defined as $X = \frac{1}{2}{\partial_{\mu}}\phi{\partial^{\mu}}\phi = \frac{1}{2}{\dot{\phi}}^2$ for a spatially homogeneous scalar field and $S_{m}$ represents the action of the background matter field. 
\vspace{3mm}\\
Varying this action with respect to the metric $g^{\mu\nu}$ gives the Einstein field equations as
\be
R_{\mu\nu} - \frac{1}{2}g_{\mu\nu}R = \frac{\partial {\cal L}}{\partial X}{{\partial_{\mu}}\phi {\partial_{\nu}}\phi} - g_{\mu\nu}{\cal L} + T^{m}_{\mu\nu}
\ee
where $T^m_{\mu\nu}$ represents the energy-momentum tensor of the matter field which is modeled in the form of an ideal perfect fluid and is defined as
\be
T^m_{\mu\nu} = (\rho_{m} + p_{m})u_{\mu}u_{\nu} - p_{m}g_{\mu\nu}
\ee 
where ${\rho}_{m}$ is the energy density, $p_{m}$ is pressure of the matter field respectively and $u_{\mu}$ is the four-velocity of the fluid. Secondly, variation of the action with respect to the scalar field $\phi$ gives the equation of motion for $\phi$ as
\be
{\ddot{\phi}}\left(\frac{\partial {\cal L}}{\partial X} + 2X\frac{\partial^2 {\cal L}}{\partial X^2}\right) + \left(3H\frac{\partial {\cal L}}{\partial X} + {\dot{\phi}}\frac{\partial^2 {\cal L}}{\partial X\partial \phi}\right){\dot{\phi}} - \frac{\partial {\cal L}}{\partial \phi} = 0
\ee
The energy density $(\rho_{\phi})$ and the pressure $(p_{\phi})$ of such a field is given by
\be\label{rhopphi}
\rho_{\phi} = \left(\frac{\partial {\cal L}}{\partial X}\right)2X - {\cal L}, \hspace{5mm} p_{\phi} = {\cal L}
\ee
In general, the Lagrangian density for a scalar field can be written as \cite{melchiorri} 
\be\label{generalL}
{\cal L}(\phi, X) = f(\phi)F(X) - V(\phi)
\ee
where $V(\phi)$ is a self-interacting potential for the scalar field $\phi$, $F(X)$ is an arbitrary function of $X$. When $f(\phi)= 1$ and $F(X)=X$ the Lagrangian (\ref{generalL}) reduces to the quintessence Lagrangian. It describes k-essence when $V(\phi)=0$ and phantom scalar field when $f(\phi)=1$ and $F(X)=-X$. In case of the phantom field, the sign of its kinetic term $``X"$ is opposite compared to the action for a canonical scalar field.\\ 
\par The Lagrangian density for a general non-canonical scalar field is given by \cite{Fang}
\be\label{ldensity}
{\cal L}(\phi, X) = F(X) - V(\phi)
\ee
These type of scalar field models with non-canonical kinetic term have received huge attention recently. Unnikrishnan et al. \cite{unni} have showed that for such models, the slow-roll conditions can be more easily satisfied compared to the canonical case. The non-canonical scalar field models are also found to be able to generate inflation in the early epoch \cite{ncm2,ncm2gv,ncm1,ncm1rj,ncm1db,ncm23,ncm231,ncm232,ncm233,ncm3,ncm301,ncm302,
unni,tgolan,Fang}. These attractive features of a noncanonical scalar field motivated us to study the features of this model in the context of dark energy. 
In this present paper, we have considered a Lagrangian density of the following form 
\be\label{L}
{\cal L}(\phi,X) = X^2 - V(\phi),\hspace{5mm}X = \frac{1}{2}{\dot{\phi}}^{2}
\ee
which can be derived from the general form of Lagrangian density \cite{unni, unni1, mukhanov} 
\be
{\cal L}(\phi, X) = X{\left(\frac{X}{M^{4}_{Pl}}\right)}^{\alpha^{\prime} -1} - V(\phi)
\ee
for $\alpha^{\prime}=2$ and $M_{Pl}=\frac{1}{\sqrt{8\pi G}}=1$.\\
It must be noted that the above equation reduces to the well known Lagrangian density for a canonical scalar field model when $\alpha^{\prime}=1$. It is also worth mentioning that the Lagrangian (\ref{L}) differs from the Lagrangian of k-essence and phantom models in the sense that for phantom field kinetic energy term itself is negative and here the potential term is non-zero. This type of  Lagrangian has also been considered in our earlier work \cite{nc}. In the subsequent sections, we try to build up an accelerating model for the universe in which the non-canonical scalar field will play the role of dynamical dark energy. We are basically interested to study how the dynamics of the non-canonical dark energy model gets affected by the various parametrizations of DE equation of state parameter. The behaviour of canonical scalar field models are very well studied for different parametrizations of DE EoS. However, the effect of these parametrizations have not been studied for a non-canonical scalar field sector.\\
\par The energy density and pressure associated with this Lagrangian density can be obtained from equations (\ref{rhopphi}) and (\ref{L}) as
\be\label{rhophi}
\rho_{\phi} = \frac{3}{4}{\dot{\phi}}^4 + V(\phi)
\ee
\be\label{pphi}
p_{\phi} = \frac{1}{4}{\dot{\phi}}^4 - V(\phi)
\ee
The metric for a homogeneous, isotropic and spatially flat FRW model of the universe is characterized by the following line element
\be\label{metric}
ds^{2} = dt^{2} - a^{2}(t)[dr^{2} + r^{2}d{\theta}^{2} +r^{2}sin^{2}\theta d{\phi}^{2}]
\ee
where $a{\rm{(t)}}$ is the scale factor, normalized so that at present $a{\rm{(t)}|_{t=t_{0}}} = 1$ and $t$ is the cosmic time. The Einstein field equations for the space-time given by equation (\ref{metric}) with matter in the form of pressureless perfect fluid takes the form, 
\be\label{eq1}
3H^{2} = {\rho}_{m} + \frac{3}{4}{\dot{\phi}}^4 + V(\phi)
\ee
\be\label{eq2}
2{\dot{H}} + 3H^{2} = -\frac{1}{4}{\dot{\phi}}^4 + V(\phi)
\ee
\be\label{eq3}
{\dot{\rho}}_{\phi} + 3H(\rho_{\phi} + p_{\phi}) = 0
\ee
\be\label{eq4}
{\dot{\rho}}_{m} + 3H{\rho}_{m} = 0
\ee
Here an overdot indicates differentiation with respect to the cosmic time $t$. Among the above four equations (equations (\ref{eq1})-(\ref{eq4})), only three are independent equations with four unknown parameters $H$, $\rho_{m}$, $\phi$ and $V(\phi)$. So we still have freedom to choose one parameter to close the above system of equations. For the present work, we consider various functional forms of the EoS parameter $\omega_{\phi}$ for the scalar field.\\
\par It is well known that the parametrization of DE equation of state plays an crucial role in understanding the nature of DE component. In general, the EoS parameter can be parametrized as,
\be
\omega_{\phi}(z) = \frac{p_{\phi}}{\rho_{\phi}}=\omega_{0} + \omega_{1}f(z)
\ee
where $\omega_{0}$, $\omega_{1}$ are real numbers and $f(z)$ is a function of redshift $z$. It may be noted that the standard flat $\Lambda$CDM model is represented by this parametrization with the choice of $\omega_{0}=-1$ and $\omega_{1}=0$. In fact, many functional forms of $f(z)$ have been considered in literature \cite{cpl0,cpl01,cpl1,cpl2,cpl3,ba,cpl5,cpl6,cpl7,cpl10,cpl20}. In this present work, we shall use four popular parametrizations of $\omega_{\phi}(z)$ to study the behavior of the deceleration parameter $q(z)$ of this non-canonical scalar field model. 
\subsection{Chevallier-Polarski-Linder (CPL) parametrization} 
Among various parametrizations, the CPL parametrization (for details, see Refs.\cite{cpl10,cpl20}) is one of the most popular ones and is given by 
\be\label{cpl}
\omega_{\phi}(z) = \omega_{0} + \omega_{1}(1 - a) = \omega_{0} + \omega_{1}\left(\frac{z}{1+z}\right)
\ee
where $z = \frac{1}{a} - 1$ is the redshift,  $\omega_{0}$  represents the current value of $\omega_{\phi}(z)$ and the second term accounts for the variation of the EoS parameter with respect to redshift. In this present model, we have considered CPL parametrization of EoS parameter because this parametrization has the advantage of giving finite $\omega_{\phi}$ in the entire range, $0 < z < \infty$.\\  
The solution for $\rho_{\phi}$ from equation (\ref{eq3}) is obtained as
\be
\rho_{\phi}(z) = \rho_{\phi 0} (1+z)^{3\alpha_{1}} e^{\left(-\frac{3\omega_{1} z}{1+z}\right)}
\ee
where, $\alpha_{1} = (1 + \omega_{0} + \omega_{1})$ and $\rho_{\phi 0}$ is an integrating constant. From equation (\ref{eq4}), we have the expression for energy density of matter as
\be
\rho_{m}(z) = \rho_{m0} (1+z)^{3}
\ee
where, $\rho_{m0}$ is an integrating constant. From equation (\ref{eq1}), the Hubble expansion rate can also be written as
\be
H^2(z) = H^2_{0}\left[\Omega_{m0} (1+z)^{3} + \Omega_{\phi 0} (1+z)^{3\alpha_{1}} e^{\left(-\frac{3\omega_{1} z}{1+z}\right)}\right]
\ee 
Here, $H_{0}$ is the Hubble parameter at the present epoch, $\Omega_{m0} = \frac{\rho_{m0}}{3H^2_{0}}$ and $\Omega_{\phi 0} = \frac{\rho_{\phi 0}}{3H^2_{0}}$ are the density parameters at the present epoch of the matter and scalar field respectively.\\
\par The deceleration parameter is defined as, $q = -\frac{{\ddot{a}}}{aH^2} = -(1 + \frac{\dot{H}}{H^2})$. For this model, $q$ takes the following form   
\be
q(z) = \frac{1}{2} + \frac{3}{2}\left[\frac{\omega_{0} + \omega_{1}\left(\frac{z}{1+z}\right)}{1 + \kappa (1 + z)^{(3 - 3\alpha_{1})}  e^{\left(\frac{3\omega_{1}z}{1+z}\right)}}\right]
\ee
where, $\kappa = \frac{\rho_{m0}}{\rho_{\phi 0}} = \frac{\Omega_{m0}}{\Omega_{\phi 0}}$. $q<0$ indicates accelerated phase of the universe while $q>0$ indicates a decelerated phase of expansion. From figure (\ref{figq}a), we see that $q$ decreases from positive to negative value for suitable choices of model parameters.
\begin{figure}[!h]
\centerline{\psfig{figure=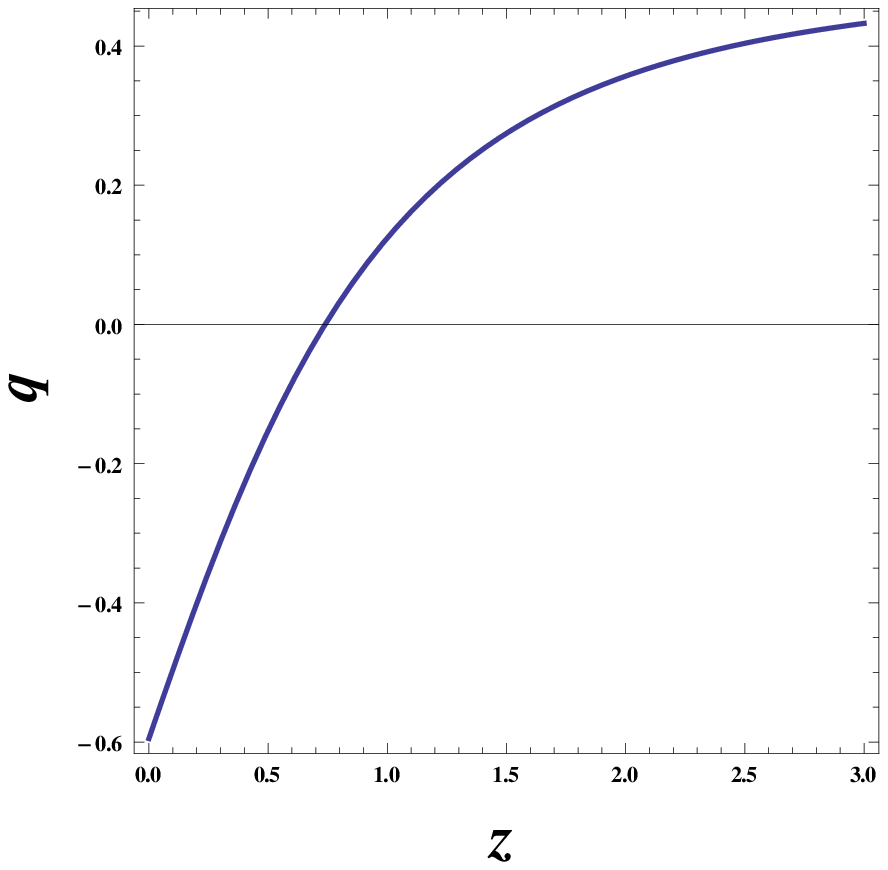,height=60mm,width=60mm}\hspace{5mm}\psfig{figure=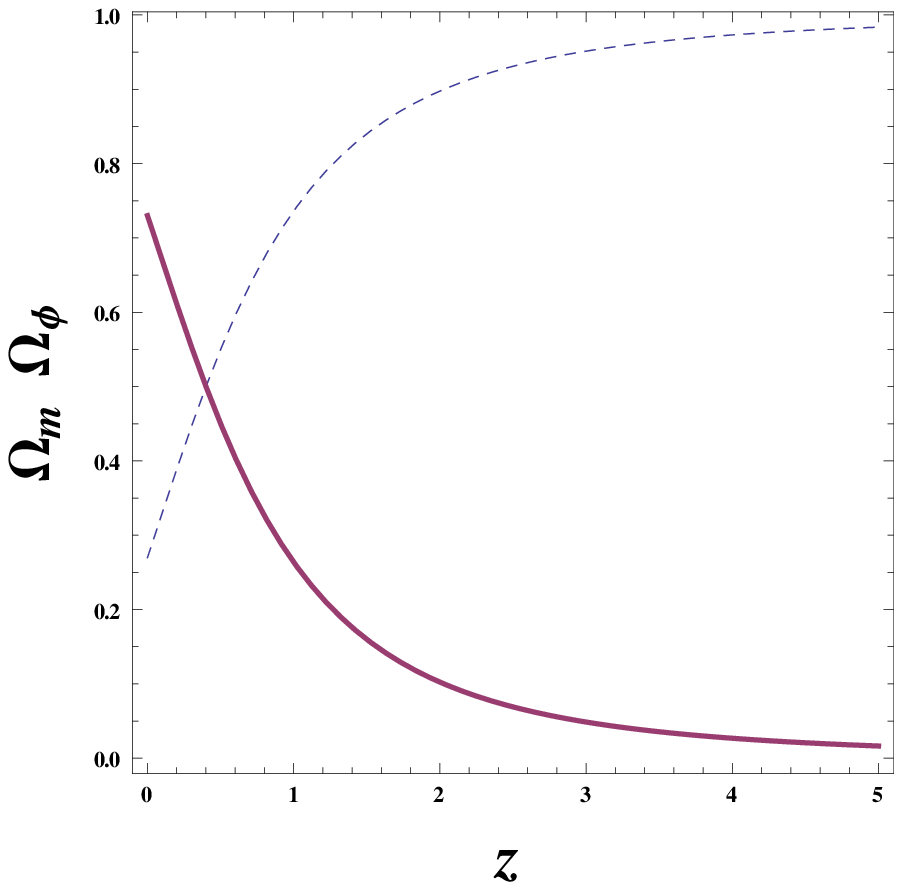,height=60mm,width=60mm}}
\caption{\normalsize{\em a) Plot of $q$ as a function of $z$ (upper panel) and b) Plot of $\Omega_{m}$ (dashed curve) and $\Omega_{\phi}$ (solid curve) as a function of $z$ (lower panel). This is for $\kappa = \frac{\Omega_{m0}}{\Omega_{\phi 0}} = \frac{0.27}{0.73}$, $\omega_{0} = -1$ and $\omega_{1} = 0.1$. Here, $\alpha_{1} = (1 + \omega_{0} + \omega_{1})$.}}
\label{figq}
\end{figure}\\
\par For this model, the evolution of the density parameters of the matter and scalar field are obtained respectively as,
\be
\Omega_{m}(z) = \frac{1}{1 + \frac{1}{\kappa} (1+z)^{3\alpha_{1} - 3} e^{\left(-\frac{3\omega_{1}z}{1+z}\right)}}
\ee
\be
\Omega_{\phi}(z) = \frac{1}{1 + \kappa (1+z)^{3 - 3\alpha_{1}} e^{\left(\frac{3\omega_{1}z}{1+z}\right)}}
\ee
which further yields, $\Omega_{m}(z) + \Omega_{\phi}(z) = 1$.
Figure (\ref{figq}b) shows the plot of density parameters for the scalar and the matter field as a function of $z$. This graph shows that $\Omega_{\phi}$ starts dominating over $\Omega_{m}$ at around $z\sim 0.4$. This result is compatible with the observational results \cite{omegaz,omegaz01}.\\
\par However, the model presented here is restricted because $\omega_{\phi}(z)$ diverges when $z\rightarrow -1$ i.e., this model cannot predict about the future evolution. So, this particular toy model is capable of describing the evolution history of the universe from the past to the near future upto $z > -1$ but can not predict about the evolution beyond that limit.  
\subsection{Jassal-Bagla-Padmanabhan (JBP) parametrization }
Recently, Jassal et al. {\cite{jbp}} extended the above parametrization to a more general case:
\be\label{eqnjbppara}
\omega_{\phi}(z) = \omega_{0} + \omega_{1}\frac{z}{(1+z)^{p}}
\ee
For the present model, we choose $p=2$. It must be noted that the EoS parameter $\omega_{\phi}\sim \omega_{0}$ at both high and low redshifts for $p=2$. Also, one can obtain the widely used CPL parametrization of EoS from equation (\ref{eqnjbppara}) for $p = 1$. For the JBP parametrization, using equation (\ref{eq3}), the expression for $\rho_{\phi}$  can be obtained as
\be
\rho_{\phi}(z) = \rho_{\phi 0}(1+z)^{3(1 + \omega_{0})} e^{\left(\frac{3\omega_{1} z^2}{2(1+z)^2}\right)} 
\ee
where $\rho_{\phi 0}$ is an integrating constant and represents the present value of the scalar field density. The Hubble parameter for this model takes the following form
\be
H^2(z) = H^2_{0}\left[\Omega_{m0} (1+z)^{3} + \Omega_{\phi 0} (1+z)^{3(1 + \omega_{0})} e^{\left(\frac{3\omega_{1} z^2}{2(1+z)^2}\right)}\right]
\ee 
In this model, we express deceleration parameter $q$ as
\be
q(z) = \frac{1}{2} + \frac{3}{2}\left[\frac{\omega_{0} + \omega_{1}\frac{z}{(1+z)^2}}{1 + \kappa (1 + z)^{- 3\omega_{0}}  e^{-\left(\frac{3\omega_{1} z^2}{2(1+z)^2}\right)}}\right]
\ee 
\begin{figure}[!h]
\centerline{\psfig{figure=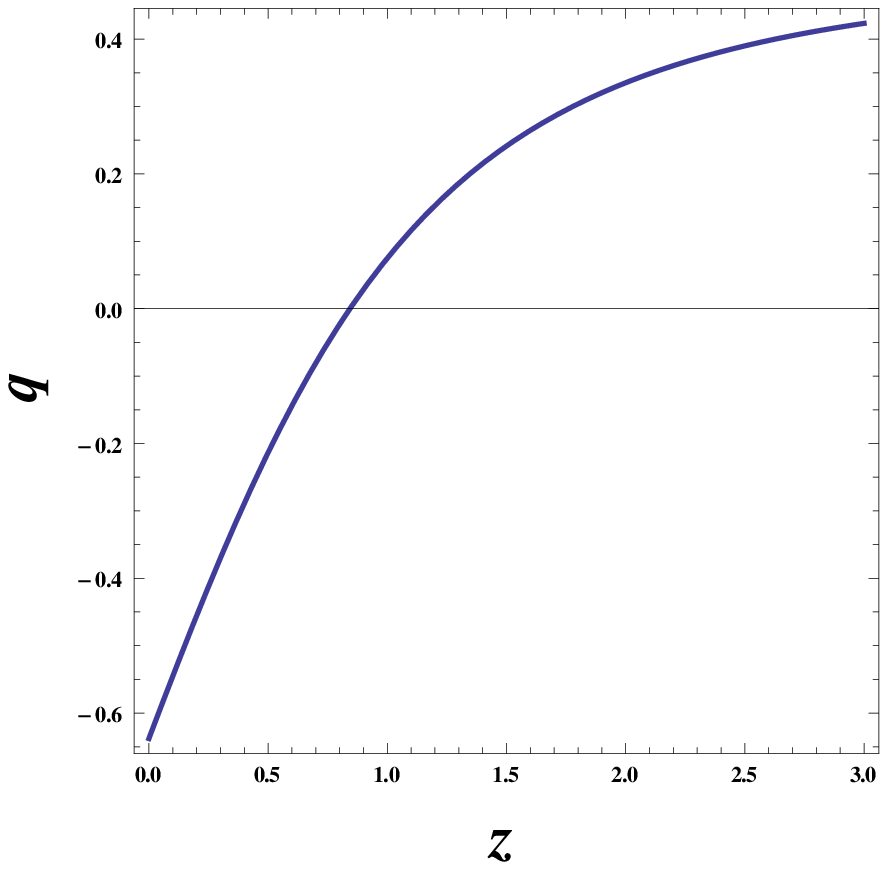,height=60mm,width=60mm}\hspace{5mm}\psfig{figure=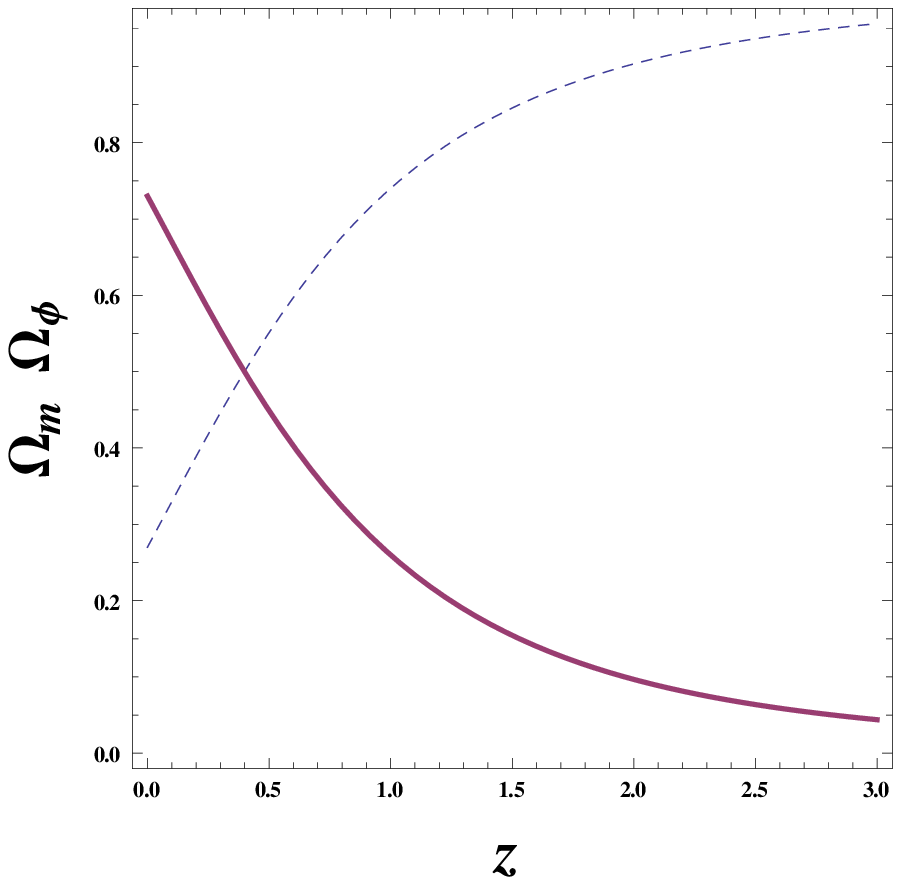,height=60mm,width=60mm}}
\caption{\normalsize{\em a) Plot of $q$ as a function of $z$ (upper panel) and b) Plot of $\Omega_{m}$ (dashed curve) and $\Omega_{\phi}$ (solid curve) as a function of $z$ (lower panel). Both the plots are for $\kappa = \frac{\Omega_{m0}}{\Omega_{\phi 0}} = \frac{0.27}{0.73}$, $\omega_{0} = -1$ and $\omega_{1} = 0.1$.}}
\label{figq1}
\end{figure}
The corresponding density parameters are now given by
\be
\Omega_{m}(z) = \frac{1}{1 + \frac{1}{\kappa} (1+z)^{3\omega_{0}} e^{\left(\frac{3\omega_{1} z^2}{2(1+z)^2}\right)}}
\ee
\be
\Omega_{\phi}(z) = \frac{1}{1 + \kappa(1+z)^{-3\omega_{0}}e^{-\left(\frac{3\omega_{1} z^2}{2(1+z)^2}\right)}}
\ee 
Figure (\ref{figq1}a) shows the plot of $q(z)$ as a function of $z$. This plot clearly shows the transition of $q$ from the decelerating to the accelerating regime at $z \sim 0.8$. The evolutions of $\Omega_{m}$ and $\Omega_{\phi}$ against $z$ are shown in figure (\ref{figq1}b). The plots are for $\omega_{0} = -1$ and $\omega_{1} = 0.1$. In both the graphs, the resulting cosmological scenarios are in good agreement with observations. For the JBP model also, $\omega_{\phi}(z)$ diverges as $z\rightarrow-1$ and thus future evolution can not be predicted.
\subsection{Barboza-Alcaniz (BA) parametrization}
The next parametrization considered in this paper was proposed by Barboza et al. \cite{ba}, which has the following functional form
\be
\omega_{\phi}(z) = \omega_{0} + \omega_{1}\frac{z(1+z)}{1+z^2}
\ee
where $\omega_{\phi}(z=0)=\omega_{0}$ (the present value of the EoS parameter), $\omega_{1} = \frac{d\omega_{\phi}}{dz}|_{z=0}$ (which measures the variation of the EoS parameter with $z$), $\omega_{\phi}(z=\infty) = \omega_{0} + \omega_{1}$ and the EoS parameter reduces to $\omega_{\phi}(z) = \omega_{0} + \omega_{1}z$ at the low redshift ($z<<1$). It is remarkable that the BA parametrization does not diverge like CPL model when $z\rightarrow -1$.\\
\par In this model, $\rho_{\phi}(z)$ becomes
\be
\rho_{\phi}(z) = \rho_{\phi 0}(1+z)^{3(1+\omega_{0})}(1+z^2)^{\frac{3\omega_{1}}{2}}
\ee
Now the equation (\ref{eq1}) can be written as
\be
H^2(z) = H^2_{0}\left[\Omega_{m0} (1+z)^{3} + \Omega_{\phi 0} (1+z)^{3(1 + \omega_{0})}(1+z^2)^{\frac{3\omega_{1}}{2}} \right]
\ee
In this case, the deceleration parameter $q(z)$ can be expressed as
\be
q(z) = \frac{1}{2} + \frac{3}{2}\left[\frac{\omega_{0} + \omega_{1}\frac{z(1+z)}{1+z^2}}{1 + \kappa (1 + z)^{- 3\omega_{0}}(1+z^2)^{-{\frac{3\omega_{1}}{2}}}}\right]
\ee
Furthermore, one can express the density parameters of the matter and scalar field respectively as
\be
\Omega_{m}(z) = \frac{1}{1 + \frac{1}{\kappa} (1+z)^{3\omega_{0}}(1+z^2)^{\frac{3\omega_{1}}{2}} }
\ee
\be
\Omega_{\phi}(z) = \frac{1}{1 + \kappa (1+z)^{-3\omega_{0}}(1+z^2)^{-\frac{3\omega_{1}}{2}} }
\ee
Figure (\ref{figq2}a) shows the evolution of the deceleration parameter with redshift $z$. It is evident from figure (\ref{figq2}a) that the universe is presently undergoing an accelerating phase of expansion ($q<0$). Figure (\ref{figq2}b) shows that the density parameter $\Omega_{m}$ increases with $z$, whereas $\Omega_{\phi}$ decreases with $z$. This features of $q(z)$, $\Omega_{m}$ and $\Omega_{\phi}$ are consistent with the present day observations.
\begin{figure}[!h]
\centerline{\psfig{figure=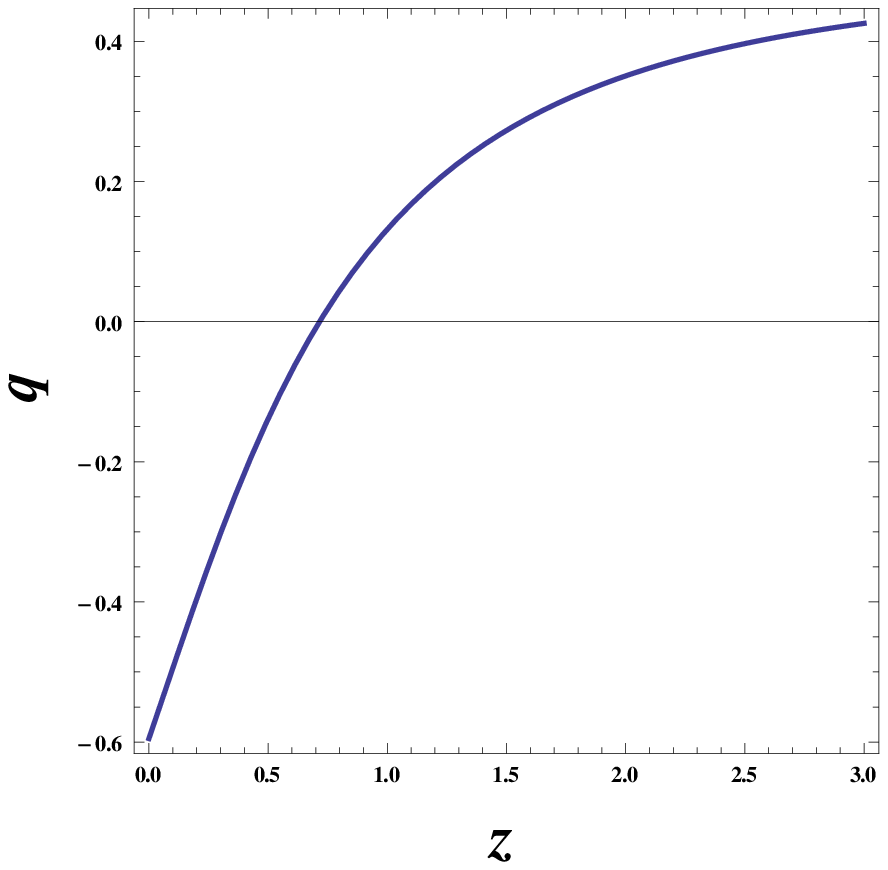,height=60mm,width=60mm}\hspace{5mm}\psfig{figure=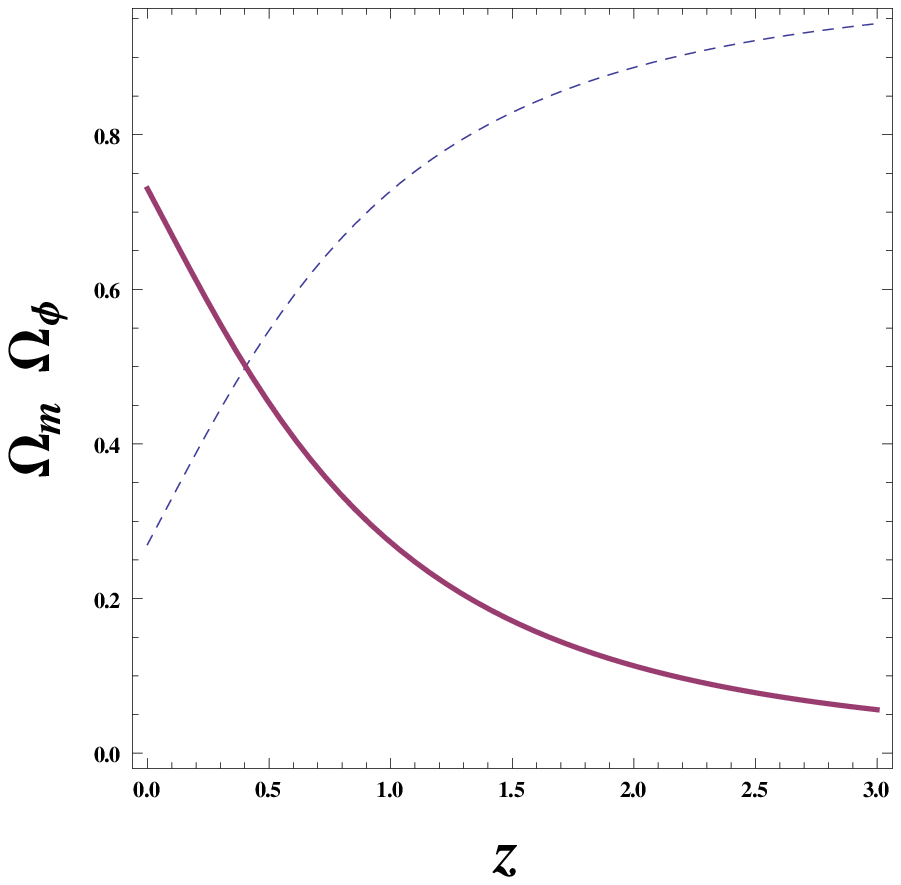,height=60mm,width=60mm}}
\caption{\normalsize{\em a) Plot of $q$ against $z$ (upper panel) and b) Plot of $\Omega_{m}$ (dashed curve) and $\Omega_{\phi}$ (solid curve) against $z$ (lower panel). Both the plots are for $\kappa = \frac{\Omega_{m0}}{\Omega_{\phi 0}} = \frac{0.27}{0.73}$, $\omega_{0} = -1$ and $\omega_{1} = 0.1$.}}
\label{figq2}
\end{figure}
\subsection{Generalized Chaplygin Gas (GCG) parametrization} 
It is well known that the generalized chaplygin gas (see Refs. \cite{cgas1,cgas2,cgas3}) behaves like dark matter in the past and it behaves like cosmological constant at present. Motivated by this idea, in this paper, we are interested to describe the late-time dynamics of the universe produced by the GCG. For this purpose, we have assumed that the universe contains both the dark matter and the GCG. Additionally, we have also considered another interesting possibility where the non-canonical scalar field $\phi$ plays the role of GCG to explore the late time cosmic scenarios. The GCG equation of state is described by \cite{cgas1,cgas2,cgas3}
\be\label{eqgcg}
p_{\phi} = - \frac{A}{\rho^{\alpha}_{\phi}}
\ee
where $A$ is a positive constant and $\alpha$ is another constant in the range $0< \alpha \leq 1$. The original chaplygin gas corresponds to the case $\alpha = 1$ \cite{cgas1}. By inserting equation (\ref{eqgcg}) into the energy conservation equation (\ref{eq3}), one finds that the density  of the scalar field $\phi$ evolves as
\be\label{rhogcg}
\rho_{\phi}(z) = {\left[A + B(1+z)^{3(1+\alpha)}\right]}^{\frac{1}{(1+\alpha)}}
\ee
where, $B$ is an integration constant. Equation (\ref{rhogcg}) can be re-written in the following form
\be
\rho_{\phi}(z) = \rho_{\phi 0}{\left[A_{s} + (1-A_{s})(1+z)^{3(1+\alpha)}\right]}^{\frac{1}{(1+\alpha)}}
\ee
where, for simplicity, we have defined $A_{s}=\frac{A}{A+B}$ and $\rho_{\phi 0} = {\left(A + B\right)}^\frac{1}{(1+\alpha)}$ is the present value of the energy density of the GCG. To ensure the finite and positive value of $\rho_{\phi}$ we need $-1< \alpha \leq 1$ and $0\le A_{s}\leq 1$.\\ 
\par In this case, the Hubble parameter is given by
\be
H^2 = H^2_{0}{\left[ \Omega_{m0}(1+z)^{3} + \Omega_{\phi 0}{\left(A_{s} + (1-A_{s})(1+z)^{3(1+\alpha)}\right)}^{\frac{1}{(1+\alpha)}}\right]}
\ee
The corresponding expression for the EoS parameter is given by
\be
\omega_{\phi}(z)=-\frac{A_{s}}{A_{s} + (1- A_{s})(1+z)^{3(1+\alpha)}}
\ee
Like earlier mentioned three models, the EoS parameter of the GCG also depends on two independent model parameters ($A_{s}$ and $\alpha$) along with redshift $z$. At present epoch, the above EoS parameter becomes, $\omega_{\phi}(z=0)= -A_{s}$. It is interesting to note that the GCG will behave like pure cosmological constant when we put $A_{s}=1$.\\
\par The deceleration parameter $q$ can be written as
\be
q(z) = \frac{1}{2} + \frac{3}{2}{\left[-\frac{\frac{A_{s}}{A_{s} + (1- A_{s})(1+z)^{3(1+\alpha)}}}{1+\frac{\kappa(1+z)^3}{{\left[A_{s} + (1-A_{s})(1+z)^{3(1+\alpha)}\right]}^{\frac{1}{(1+\alpha)}}}}\right]}
\ee
where $\kappa = \frac{\rho_{m0}}{\rho_{\phi 0}} = \frac{\Omega_{m0}}{\Omega_{\phi 0}}$.\\
\begin{figure}[!h]
\centerline{\psfig{figure=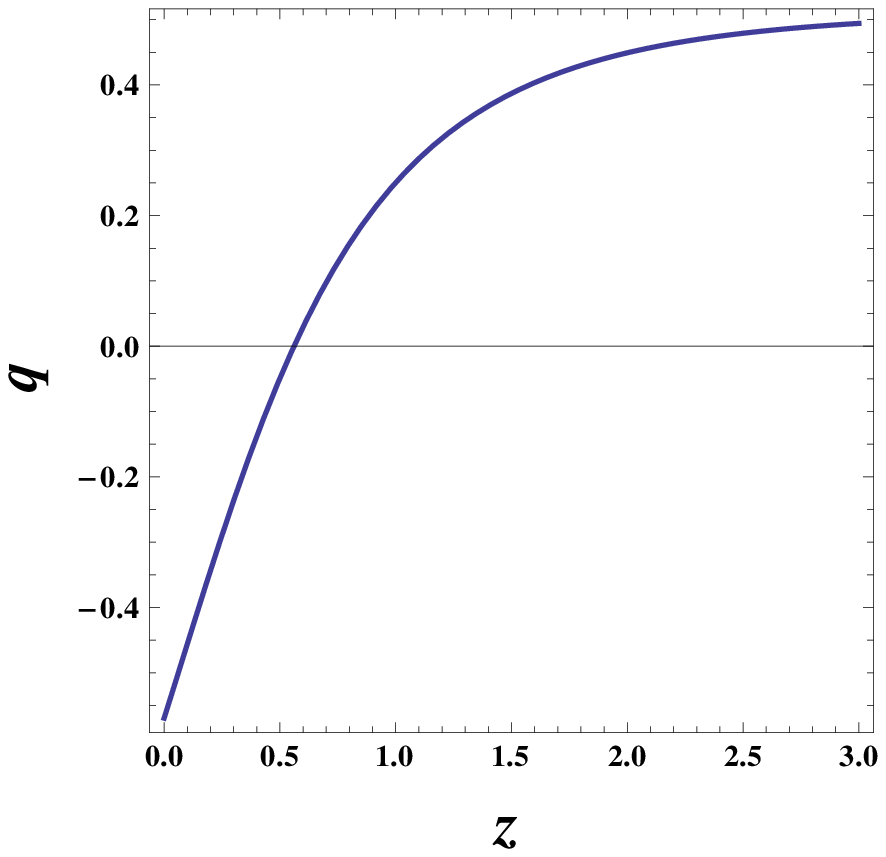,height=60mm,width=60mm}\hspace{5mm}\psfig{figure=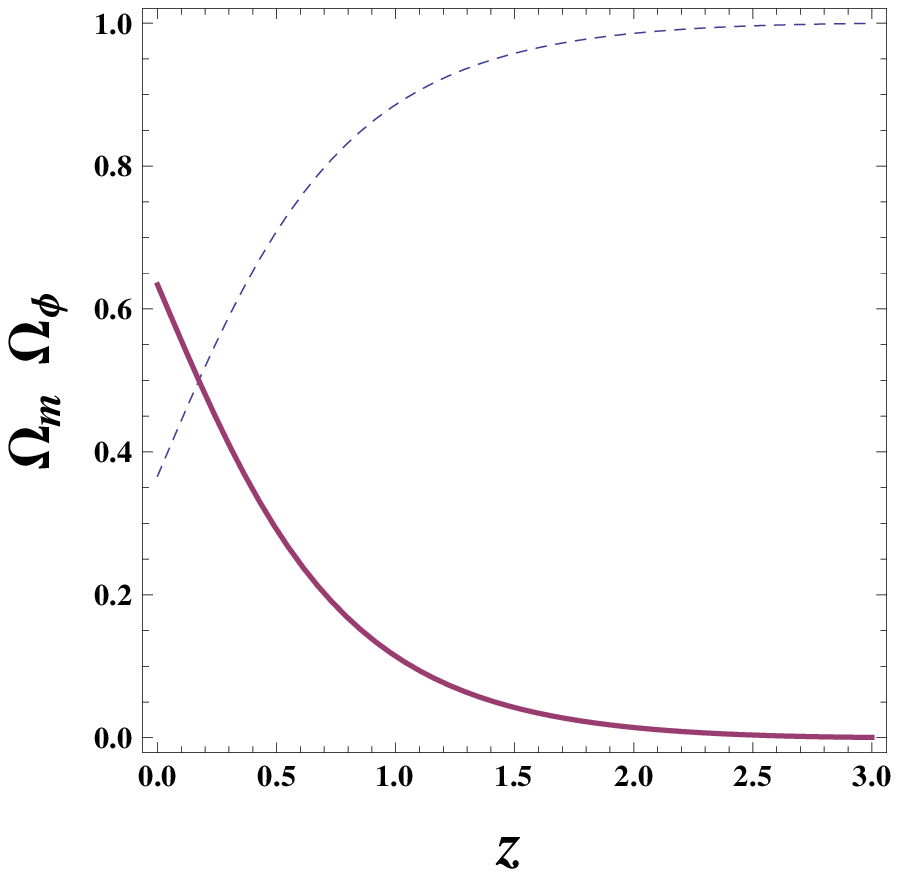,height=60mm,width=60mm}}
\caption{\normalsize{\em a) Plot of $q$ as a function of $z$ (upper panel) and b) Plot of $\Omega_{m}$ (dashed curve) and $\Omega_{\phi}$ (solid curve) as a function of $z$ (lower panel). Both the plots are for $A_{s} = 0.9$, $\alpha = -0.5$ and $\kappa=\frac{0.27}{0.73}$. }}
\label{figq3}
\end{figure}
In this case the density parameters have the following form\\
\bea
\Omega_{m}(z) = \frac{1}{1+ \frac{{\left[A_{s} + (1-A_{s})(1+z)^{3(1+\alpha)}\right]}^{\frac{1}{(1+\alpha)}}}{\kappa(1+z)^3}}\\
\Omega_{\phi}(z) = \frac{1}{{1+\frac{\kappa(1+z)^3}{{\left[A_{s} + (1-A_{s})(1+z)^{3(1+\alpha)}\right]}^{\frac{1}{(1+\alpha)}}}}}                         
\eea
Figure (\ref{figq3}a) shows the evolution of $q(z)$ vs. redshift $z$ for $A_{s} = 0.9$ and $\alpha = -0.5$. In fact, at low redshift, the transition of $q(z)$ from decelerating to accelerating regime depends upon the choice of the parameters $A_{s}$ and $\alpha$. Also, the evolutions of $\Omega_{m}$ and $\Omega_{\phi}$ against $z$ are shown in figure (\ref{figq3}b) for the earlier mentioned same chosen values of $A_{s}$ and $\alpha$.
\subsection{Comparison between canonical and non-canonical scalar field models for the above parametrizations:}
\par For all these models, the relevant potential for the scalar field $\phi$ in terms of redshift $z$ can be written as (from equations (\ref{rhophi}) and (\ref{pphi}))
\be\label{vnoncano}
V(z)=\frac{1}{4}{\left(1-3\omega_{\phi}(z)\right)}\rho_{\phi}(z)
\ee
which immediately gives
\be
V_{CPL}(z) = V_{0}{\left(1-3\omega_{0}-\frac{3\omega_{1}z}{1+z}\right)}(1+z)^{3\alpha_{1}}e^{-\frac{3\omega_{1}z}{1+z}}
\ee
\be
V_{JBP}(z) = V_{0}{\left(1-3\omega_{0}-\frac{3\omega_{1}z}{(1+z)^2}\right)}(1+z)^{3(1+\omega_{0})}e^{\frac{3\omega_{1}z^2}{2(1+z)^2}}
\ee
\be
V_{BA}(z) = V_{0}{\left(1-3\omega_{0}-3\omega_{1}\frac{z(1+z)}{1+z^2}\right)}(1+z)^{3(1+\omega_{0})}{\left(1+ z^2\right)}^{\frac{3\omega_{1}}{2}}
\ee
\be
V_{GCG}(z) 
= V_{0}{\left(4A_{s} + (1-A_{s})(1+z)^{3(1+\alpha)}\right)} {\times \left[A_{s} + (1-A_{s})(1+z)^{3(1+\alpha)}\right]}^{-\frac{\alpha}{(1+\alpha)}}
\ee
where $V_{0}=\frac{3\Omega_{\phi 0}H^2_{0}}{4}$. $V_{CPL}$, $V_{JBP}$, $V_{BA}$ and $V_{GCG}$ are the potential for the CPL, JBP, BA and GCG models respectively. Here, $H_{0}$ and $\Omega_{\phi 0}$ represent the present day values for the Hubble parameter and the dark energy density parameter respectively.\\
\begin{figure}[!h]
\centerline{\psfig{figure=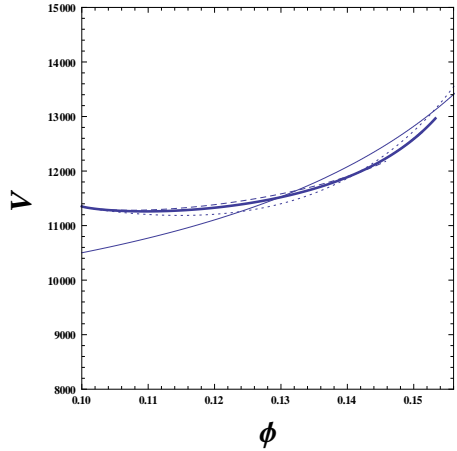,height=60mm,width=60mm}}
\caption{\normalsize{\em This figure shows the variation of the potential $V$ with $\phi$ for non-canonical scalar field model by assuming $\omega_{0}=-1$, $\omega_{1}=0.1$ for CPL (thick curve), JBP (dashed curve) and BA (dotted curve) parametrizations and $A_{s}=0.9$, $\alpha=-0.5$ for GCG (thin curve) parametrization. All the plots are for the parameter choices $\Omega_{\phi 0}=0.73$, $\phi_{0}=0.1$ and $H_{0}=72$ $km s^{-1} Mpc^{-1}$.}}
\label{figvallnc}
\end{figure}
Adding equations (\ref{rhophi}), (\ref{pphi}) and replacing ${\dot{\phi}} = aH\frac{d\phi}{da}$, one can obtain the general expression for the non-canonical scalar field $\phi$ as,
\be\label{phinoncano}
\phi(z) = \phi_{0} +\int^{z}_{0}{\frac{{\left[(1+\omega_{\phi}(z^{\prime}))\rho_{\phi}(z^{\prime})\right]}^{\frac{1}{4}}}{(1+z^{\prime})H(z^{\prime})}dz^{\prime}} 
\ee
where $\phi_{0}$ is an arbitrary integration constant.\\
\par Now, we will focus on the extensively studied canonical scalar field case, in which the Lagrangian density is obtained from equation (\ref{ldensity}) as
\be
{\cal L}(\phi_{cano}, X) = X - V(\phi)
\ee
The energy density and pressure for the scalar field $(\phi_{cano})$ are given by
\be
\rho_{cano}=\frac{1}{2}{\dot{\phi}}^2_{cano} + V_{cano}(\phi_{cano}), \hspace{3mm}p_{cano}=\frac{1}{2}{\dot{\phi}}^2_{cano} - V_{cano}(\phi_{cano})
\ee
where, $V_{cano}$ is the potential of the canonical scalar field. The EoS parameter $\omega_{cano}=\frac{p_{cano}}{\rho_{cano}}$ is a dynamical variable which gives the continuity equation (\ref{eq3}) in an integrated form
\be
\rho_{cano}=\rho_{0}exp{\left[3\int^{z}_{0}\frac{1+\omega^{\prime}_{cano}(z^{\prime})}{1+z^{\prime}}dz^{\prime}\right]}
\ee
where $\rho_{0}$ is an integration constant. The Friedmann equation then becomes
\be
H^2_{cano}(z) = H^2_{0}[\Omega_{m0} (1+z)^{3}+ (1-\Omega_{m0})exp{\left(3\int^{z}_{0}\frac{1+\omega^{\prime}_{cano}(z^{\prime})}{1+z^{\prime}}dz^{\prime}\right)}]
\ee
In this case, the expressions for $V_{cano}$ and $\phi_{cano}$ can be written as
\be\label{vcano}
V_{cano}(z)=\frac{1}{2}{\left(1-\omega_{cano}(z)\right)}\rho_{cano}(z)
\ee
and 
\be\label{phicano}
\phi_{cano}(z) = \phi_{0} +\int^{z}_{0}{\frac{{\left[(1+\omega_{cano}(z^{\prime}))\rho_{cano}(z^{\prime})\right]}^{\frac{1}{2}}}{(1+z^{\prime})H_{cano}(z^{\prime})}dz^{\prime}} 
\ee
In the previous subsection, we have discussed various parametrizations of the EoS parameter for the non-canonical scalar field. Now, we will consider these parametrizations for canonical scalar field models to compare their behavior with the non-canonical scalar field models. It deserves mention that one obtains the same expressions of $\rho_{cano}(z)$ and $H_{cano}(z)$ for both canonical and non-canonical scalar field models. But, the expressions for potential associated with the scalar field $\phi$ will be different for canonical and non-canonical scalar field models (see equations (\ref{vnoncano}), (\ref{phinoncano}), (\ref{vcano}) and (\ref{phicano})).  
\par The expressions for the potential $V(z)$ and $\phi(z)$ are very complicated and it is very difficult to express $V$ in terms of $\phi$. So, we have solved equations (\ref{phinoncano}) and (\ref{phicano}) numerically and have plotted $V$ as a function of $\phi$ (see Figures {\ref{figvallnc}} and {\ref{figv}}).
\begin{figure}[!h]
\centerline{\psfig{figure=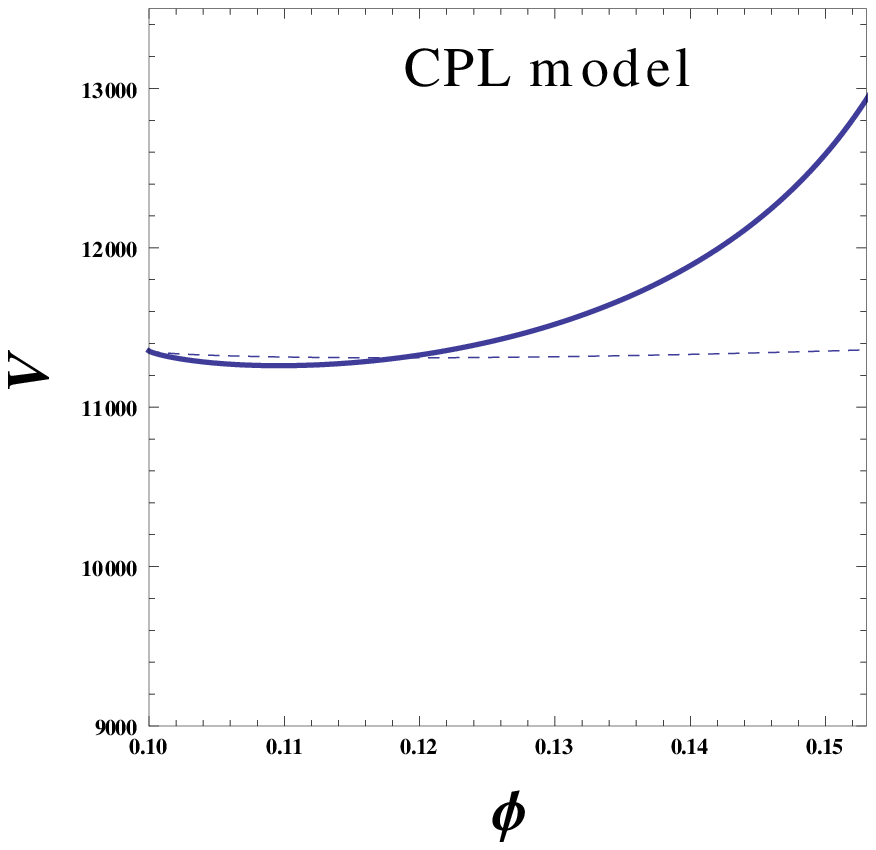,height=60mm,width=60mm}\hspace{5mm}\psfig{figure=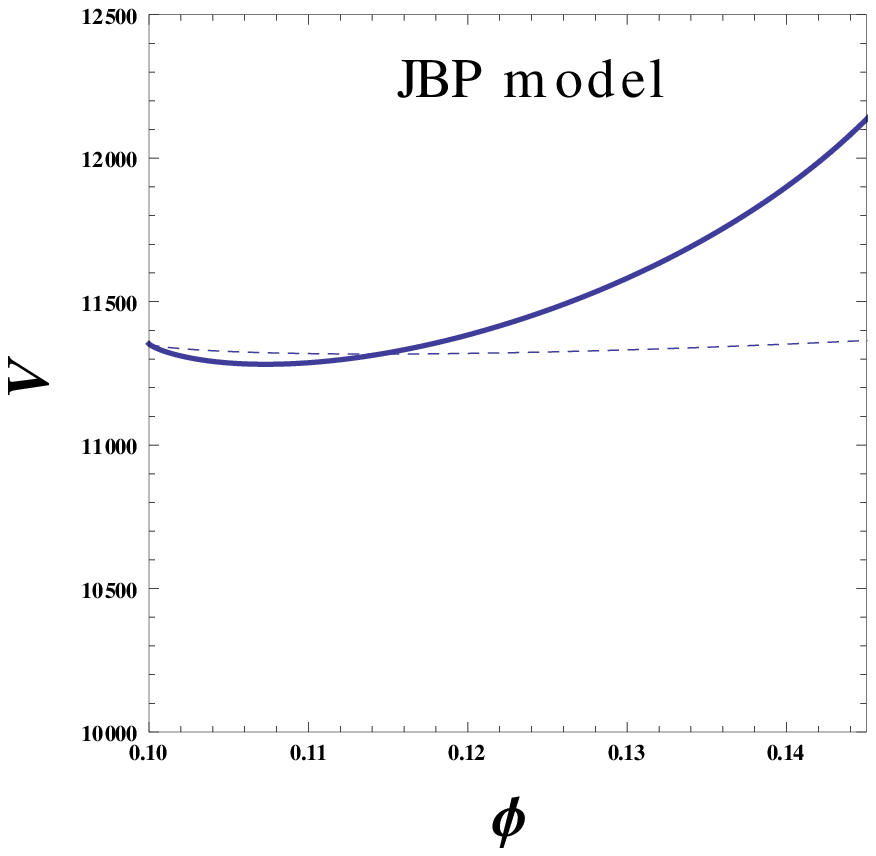,height=60mm,width=60mm}}
\centerline{\psfig{figure=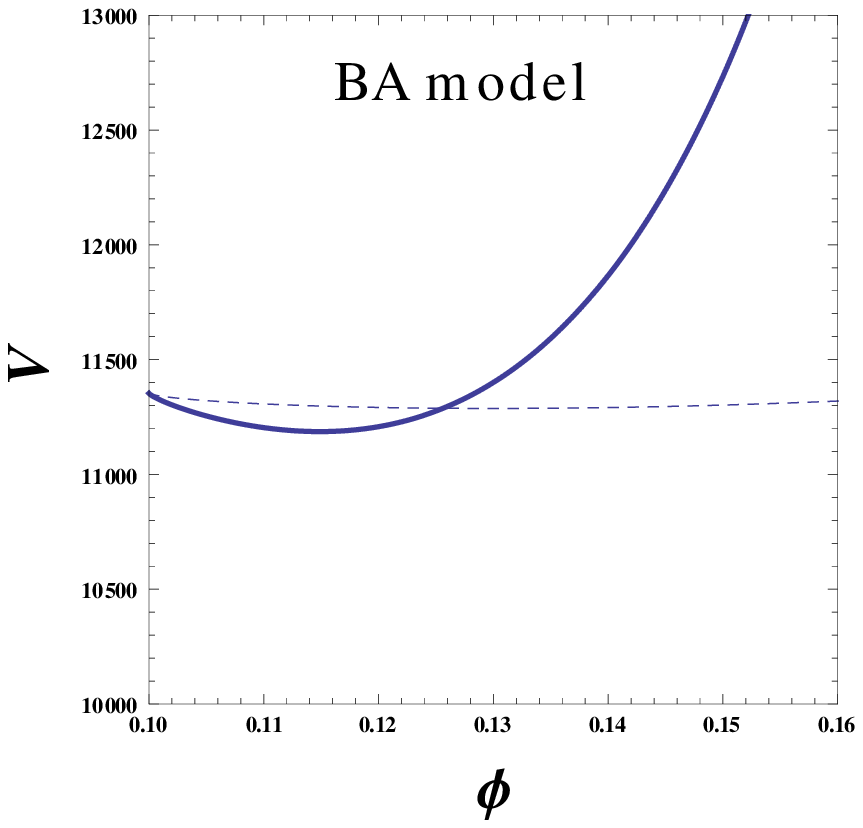,height=60mm,width=60mm}\hspace{5mm}\psfig{figure=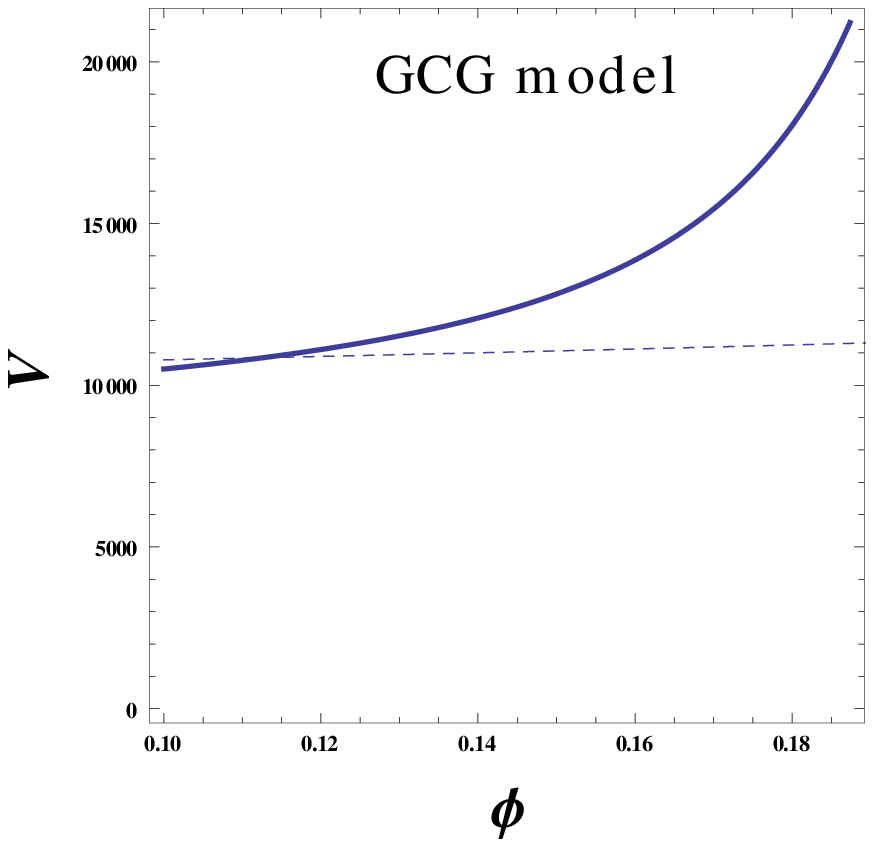,height=60mm,width=60mm}}
\caption{\normalsize{\em This figure shows the variation of the potential $V$ with $\phi$  by assuming $\omega_{0}=-1$, $\omega_{1}=0.1$ for CPL, JBP and BA parametrizations and $A_{s}=0.9$, $\alpha=-0.5$ for GCG parametrization. The dashed curve represents the trajectory of the potentials for the canonical scalar field, as shown in each panel. All the plots are for the parameter choices $\Omega_{\phi 0}=0.73$, $\Omega_{m0}=0.27$, $\phi_{0}=0.1$ and $H_{0}=72$ $km s^{-1} Mpc^{-1}$.}}
\label{figv}
\end{figure}
In figure {\ref{figv}}, for each panel, the dashed curve shows the evolution of the potential $V_{cano}(\phi_{cano})$ for each parametrization whereas the solid lines represents the evolution for corresponding non-canonical case. It has been found that for canonical case, the slope of the potential is quite flat and $\phi_{cano}$ is almost constant $({\dot{\phi}}^2_{cano}<<V_{cano})$ throughout the evolution, which yields $\rho_{cano}\approx V_{cano}=$ constant, however, the case is different for a non-canonical scalar field model. Figure {\ref{figv}} also shows for each panel, that for non-canonical case, the trajectory of the potential $V(\phi)$ changes very slowly at early epoch, but it starts increasing with $\phi$ in such way that the potential term dominates over the non-canonical kinetic term (${\dot{\phi}}^4$) at late times independent of initial conditions and thus provides acceleration. So, the non-canonical scalar field exhibits an interesting property of the potential, which can provide a possible solution to the coincidence problem. This new feature occurs due to the non-canonical kinetic term present in the Lagrangian (\ref{L}).
\section{Observational Constraints on $\omega_{\phi}(z)$}
In this section, we shall fit these four parametrized theoretical models with recent observational datasets, namely Type Ia Supernovae (SN Ia), measurements of Hubble parameter (H(z)), baryonic acoustic oscillations (BAO) and cosmic microwave background (CMB) observations. Although the procedure for calculation of individual $\chi^2$ function is quite well-known, we briefly mention the same for completeness.\\
\par The total $\chi^2$ for joint data analysis is defined as
\be
\chi^2_{total}= \chi^2_{SN} + \chi^2_{H} + \chi^2_{BAO/CMB}
\ee
where the individual $\chi^2$  for each dataset is evaluated as follows.
\par First, we have used the latest observational dataset of SN Ia {\cite{suzuki}} of 580 data points. In order to calculate $\chi^2_{SN}$ for SN Ia data we follow the procedure described in Refs. \cite{nesseris,nesseris01, Rapetti, Coe}.  
For SN Ia dataset, the $\chi^2$ function is constructed as
\be\label{chisquare} 
\chi^2_{SN} = A - \frac{B^2}{C}
\ee
where $A$, $B$ and $C$ are defined as follows
\bea
A = \sum^{580}_{i=1} \frac{[{\mu}^{obs}(z_{i}) - {\mu}^{th}(z_{i})]^2}{\sigma^2_{i}},\\
B = \sum^{580}_{i=1} \frac{[{\mu}^{obs}(z_i) - {\mu}^{th}(z_{i})]}{\sigma^2_{i}},
\eea
and
\be 
C = \sum^{580}_{i=1} \frac{1}{\sigma^2_{i}}
\ee 
where $\mu^{obs}$, $\mu^{th}$ represent the observed and theoretical distance modulus respectively and $\sigma_{i}$ represent the uncertainty in the distance modulus.\\
\par Next, we have used the 29 data points of $H(z)$ \cite{zhangh,simon, moresco,chuang,blakeh, stern,lsamu,delubac,xding} in the redshift range $0.07\le z\le 2.34$. These values are presented in table (\ref{table0}). To complete the dataset, we fixed the value of $H_{0}$ from \cite{h0}. For this dataset, the $\chi^2$ function is defined as
\be
\chi^2_{H} = \sum^{29}_{i=1}\frac{[{h}^{obs}(z_{i}) - {h}^{th}(z_{i})]^2}{\sigma^2_{H}(z_{i})} 
\ee
where ${h} = \frac{H(z)}{H_{0}}$ is the normalized Hubble parameter and ${\sigma_{H}}$ is the error associated with each data point. In above equation subscript $``obs"$ refers to observational quantities and subscript $``th"$ refers to the corresponding theoretical ones.\\
\par We have also used BAO \cite{baob, baop, baobl} and CMB \cite{wmap7} measurements dataset to obtain the BAO/CMB constraints on the model parameters. In order to calculate $\chi^2_{BAO/CMB}$ function, we follow the procedure described in Ref. \cite{goistri}. We use the measurement of ``acoustic scale" $l_{A}$ provided by CMB, which is defined as
\be
l_{A}=\pi\frac{d_{A}(z_{*})}{r_{s}(z_{*})}
\ee
where $d_{A}(z_{*})=\int^{z_{*}}_{0}\frac{dz^{\prime}}{H(z^{\prime})}$, is the comoving angular diameter distance, $r_{s}(z_{*})$ is the comoving sound horizon at the photon-decoupling epoch and $z_{*} \approx 1091$ is the decoupling time. In this work, we have used $l_{A}=302.44 \pm 0.80$ \cite{wmap7}. For this analysis, we have used six data points from 6dFGS \cite{baob}, SDSS LRG \cite{baop} and WiggelZ \cite{baobl} surveys. Following Ref. \cite{goistri} and combining these results with \cite{wmap7}, we obtained the BAO/CMB constraints on the model parameters (see table \ref{table01}).\\
The $\chi^2$ for the BAO/CMB data is given by 
\be
\chi^2_{BAO/CMB} = X^{T}C^{-1}X 
\ee
where, $X$ (transformation matrix) and $C^{-1}$ (inverse covariance matrix) which are formed using different functional form of $\frac{d_{A}(z_{*})}{D_{v}(z_{BAO})}$. For details refer to Goistri et al. \cite{goistri}.\\
\par The CPL, JBP and BA model have three free parameters, namely, $\Omega_{m0}$ (or $\Omega_{\phi 0}= 1- \Omega_{m0}$), $\omega_{0}$ and $\omega_{1}$. In this case, the confidence region ellipses in the $\omega_{0} - \omega_{1}$ parameter space can be drawn by fixing $\Omega_{m0}$ to some constant value. So, we have done $\chi^2$ analysis by fixing $\Omega_{m0}$ (the present value of the density parameter of the matter field) to $0.26$, $0.27$ and $0.28$ for those dataset. Hence, we can now deal with only two free parameters ($\omega_{0}$, $\omega_{1}$) and will obtain the observational bounds on this parameter from $\chi^2$ analysis of the combined dataset (SNIa $+$ H(z) $+$ BAO/CMB). With this we have plotted $1\sigma$ $(68.3\%)$ and $2\sigma$ $(95.4\%)$ confidence contours on  $\omega_{0} - \omega_{1}$ parameter space for various DE parametrizations. Similarly, the GCG model has three free parameters, namely, $\Omega_{m0}$, $A_{s}$ and $\alpha$. In this case, we have plotted  $1\sigma$ and $2\sigma$ confidence contours on  $A_{s} - \Omega_{m0}$ parameter space by considering $\alpha=-0.5$. The best-fit values of the parameters are obtained by minimizing  $\chi^2$. In figure (\ref{figcontour1}) and (\ref{figcontour2}), the large dots represent the best fit values of the model parameters and the small dots represent the chosen values of these parameters in our analytical models (as mentioned in previous section). In our analytical models, we have chosen $\omega_{0} = -1$, $\omega_{1} = 0.1$ (for CPL, JBP and BA model) and $A_{s}=0.9$, $\Omega_{m0}=0.27$ (for GCG model) as we are interested to study and understand the effect of individual DE parametrizations for some fixed values of model parameters. For this combined datasets, the chosen values of $\omega_{0}$ and $\omega_{1}$ ($A_{s}$ and $\alpha$) are found to be well within the $1\sigma$ confidence contour. The observational bound on these model parameters as well as their best fit values are presented in table \ref{table1} $\&$ \ref{table2}. For each model, we notice from table \ref{table1} $\&$ \ref{table2} that the best fit value of the present EoS parameter is very close to $-1$ which is consistent with the recent observations \cite{wv,davis}.  The standard flat $\Lambda$CDM model ($\omega_{0}=-1$ and $\omega_{1}=0$) corresponds to the intersection point of the dashed lines as plotted in figure (\ref{figcontour1}), and it is evident from the figures that the $\Lambda$CDM model is always
 inside the $1\sigma$ confidence contour for all these parameterizations.
\begin{table*}
\caption{$H(z)$ measurements (in unit [$km$ $s^{-1}$ $Mpc^{-1}$]) in the redshift range $0.07\le z \le 2.34$ and their errors.}
\label{table0}
\begin{tabular*}{\textwidth}{@{\extracolsep{\fill}}lrrl@{}}
\hline
$z$ & \multicolumn{1}{c}{$H(z)$} & \multicolumn{1}{c}{$\sigma_{H}$} & \multicolumn{1}{c}{Reference}\\
\hline
0.07& 69& 19.6& Zhang et al. \cite{zhangh}\\
\hline
0.1& 69& 12& Simon et al. \cite{simon}\\
\hline
0.12& 68.6& 26.2& Zhang et al. \cite{zhangh}\\
\hline
0.17& 83& 8& Simon et al. \cite{simon}\\
\hline
0.179& 75& 4& Moresco et al. \cite{moresco}\\
\hline
0.199& 75& 5& Moresco et al. \cite{moresco}\\
\hline
0.2& 72.9& 29.6& Zhang et al. \cite{zhangh}\\
\hline
0.27& 77& 14& Simon et al. \cite{simon}\\
\hline
0.28& 88.8& 36.6& Zhang et al. \cite{zhangh}\\
\hline
0.35& 82.7& 8.4& Chuang et al. \cite{chuang}\\
\hline
0.352& 83& 14& Moresco et al. \cite{moresco}\\
\hline
0.4& 95& 17& Simon et al. \cite{simon}\\
\hline
0.44& 82.6& 7.8& Blake et al. \cite{blakeh}\\
\hline
0.48& 97& 62& Stern et al. \cite{stern}\\
\hline
0.57& 92.9& 7.8& Samushia et al. \cite{lsamu}\\
\hline
0.593& 104& 13& Moresco et al. \cite{moresco}\\
\hline
0.6& 87.9& 6.1& Blake et al. \cite{blakeh}\\
\hline
0.68& 92& 8& Moresco et al. \cite{moresco}\\
\hline
0.73& 97.3& 7& Blake et al. \cite{blakeh}\\
\hline
0.781& 105& 12& Moresco et al. \cite{moresco}\\
\hline
0.875& 125& 17& Moresco et al. \cite{moresco}\\
\hline
0.88& 90& 40& Stern et al. \cite{stern}\\
\hline
0.9& 117& 23& Simon et al. \cite{simon}\\
\hline
1.037& 154& 20& Moresco et al. \cite{moresco}\\
\hline
1.3& 168& 17& Simon et al. \cite{simon}\\
\hline
1.43& 177& 18& Simon et al. \cite{simon}\\
\hline
1.53& 140& 14& Simon et al. \cite{simon}\\
\hline
1.75& 202& 40& Simon et al. \cite{simon}\\
\hline
2.34& 222& 7& Delubac et al. \cite{delubac}, Ding et al. \cite{xding}\\
\hline
\end{tabular*}
\end{table*}
\begin{table*}
\caption{Values of $\frac{d_{A}(z_{*})}{D_{V}(z_{BAO})}$ for different values of $z_{BAO}$. Here, $D_{V}={\left[d^{2}_{A}(z)\frac{z}{H(z)}\right]}^{\frac{1}{3}}$ is the dilation scale \cite{dscale}.}
\label{table01}
\begin{tabular*}{\textwidth}{@{\extracolsep{\fill}}lrl@{}}
\hline
$z_{BAO}$ & \multicolumn{1}{c}{$\frac{d_{A}(z_{*})}{D_{V}(z_{BAO})}$} & \multicolumn{1}{c}{Reference}\\
\hline
0.106& 30.95 $\pm$ 1.46&Beutler et al. \cite{baob}\\
\hline
0.2& 17.55 $\pm$ 0.60&Percival et al. \cite{baop}\\
\hline
0.35& 10.11 $\pm$ 0.37& Percival et al. \cite{baop}\\
\hline
0.44& 8.44 $\pm$ 0.67& Blake et al. \cite{baobl}\\
\hline
0.6& 6.69 $\pm$ 0.33&  Blake et al. \cite{baobl}\\
\hline
0.73& 5.45 $\pm$ 0.31&  Blake et al. \cite{baobl}\\
\hline
\end{tabular*}
\end{table*}
\begin{table*}
\caption{Best fit values of $\omega_{0}$, $\omega_{1}$ and the minimum values of $\chi^2$ corresponding to the joint analysis of SN Ia $+$ H(z) $+$ BAO/CMB dataset with different choices of $\Omega_{m0}$.}
\label{table1}
\begin{tabular*}{\textwidth}{@{\extracolsep{\fill}}lrrrrl@{}}
\hline
Name & \multicolumn{1}{c}{$\Omega_{m0}$} & \multicolumn{1}{c}{${\omega}_{0}$} & \multicolumn{1}{c}{$\omega_{1}$} & \multicolumn{1}{c}{Constraints on ${\omega}_{0}$ and $\omega_{1}$ } & $\chi^2_{min}$ \\
&&&&(within $1\sigma$ confidence level) &\\
\hline
CPL model&0.26&-1.07165&0.424314& -1.228$\leq \omega_{0}\leq$ -0.9161,&579.536\\
&&&&-0.3951$\leq \omega_{1}\leq$ 1.23 &\\
&0.27& -1.04934& 0.132091& -1.22$< \omega_{0}\leq$ -0.8766, &578.705\\
&&&&-0.7968$\leq \omega_{1}\leq$ 1.062 &\\
&0.28&-1.02831&0.103564& -1.199$\leq \omega_{0}\leq$ -0.8572,&588.908\\
&&&&-0.8023$\leq \omega_{1}\leq$ 0.9991&\\
\hline
JBP model&0.26&-1.07798&0.6&-1.301$\leq \omega_{0}\leq$ -0.8525,&579.667\\
&&&&-1.064$\leq \omega_{1}\leq$ 2.248&\\
&0.27& -1.06072&0.261937&-1.293$\leq \omega_{0}\leq$ -0.825,  &578.692 \\
&&&&-1.518$\leq \omega_{1}\leq$ 2.003 &\\
&0.28&-1.05974&0.024530&-1.292$\leq \omega_{0}\leq$ -0.8242,&578.224\\
&&&&-1.789$\leq \omega_{1}\leq$ 1.737 &\\
\hline
BA model&0.26&-1.05344&0.202546&-1.182$\leq \omega_{0}\leq$ -0.9232&579.612\\
&&&&-0.2033$\leq \omega_{1}\leq$ 0.6094&\\
&0.27& -1.04272 & 0.059557&-1.182$\leq \omega_{0}\leq$ -0.9043, &578.712\\
&&&&-0.4118$\leq \omega_{1}\leq$ 0.5235 &\\
&0.28&-1.06739&0.018632&-1.213$\leq \omega_{0}\leq$ -0.9202,&578.276\\
&&&&-0.4808$\leq \omega_{1}\leq$ 0.515&\\
\hline
\end{tabular*}
\end{table*}
\begin{table*}
\caption{Best fit values of $A_{s}$, $\Omega_{m0}$ and the minimum values of $\chi^2$ corresponding to the joint analysis of SN Ia $+$ H(z) $+$ BAO/CMB dataset with different values of $\alpha$. Note that $\omega_{\phi}(z=0)= -A_{s}$ is the present day value of the EoS parameter for the GCG model.}
\label{table2}
\begin{tabular*}{\textwidth}{@{\extracolsep{\fill}}lrrrl@{}}
\hline
$\alpha$ & \multicolumn{1}{c}{$A_{s}$} & \multicolumn{1}{c}{$\Omega_{m0}$} & \multicolumn{1}{c}{Constraints on $A_{s}$ and $\Omega_{m0}$ } & \multicolumn{1}{c}{$\chi^2_{min}$}\\
&&&(within $1\sigma$ confidence level)&\\
\hline
-0.5& 0.9& 0.240784& 0.8979$\leq A_{s}\leq$ 0.9462, &585.915\\
&&& 0.20$\leq \Omega_{m0}\leq$0.3396 &\\
\hline
-0.55& 0.9& 0.243458& 0.8976$\leq A_{s}\leq$ 0.9467, &586.196\\
&&& 0.20$\leq \Omega_{m0}\leq$0.3432 &\\
\hline
-0.6& 0.9& 0.245692& 0.8997$\leq A_{s}\leq$ 0.9446, &586.344\\
&&& 0.20$\leq \Omega_{m0}\leq$0.3459 &\\
\hline
\end{tabular*}
\end{table*}
\begin{figure}[!h]
\centerline{\psfig{figure=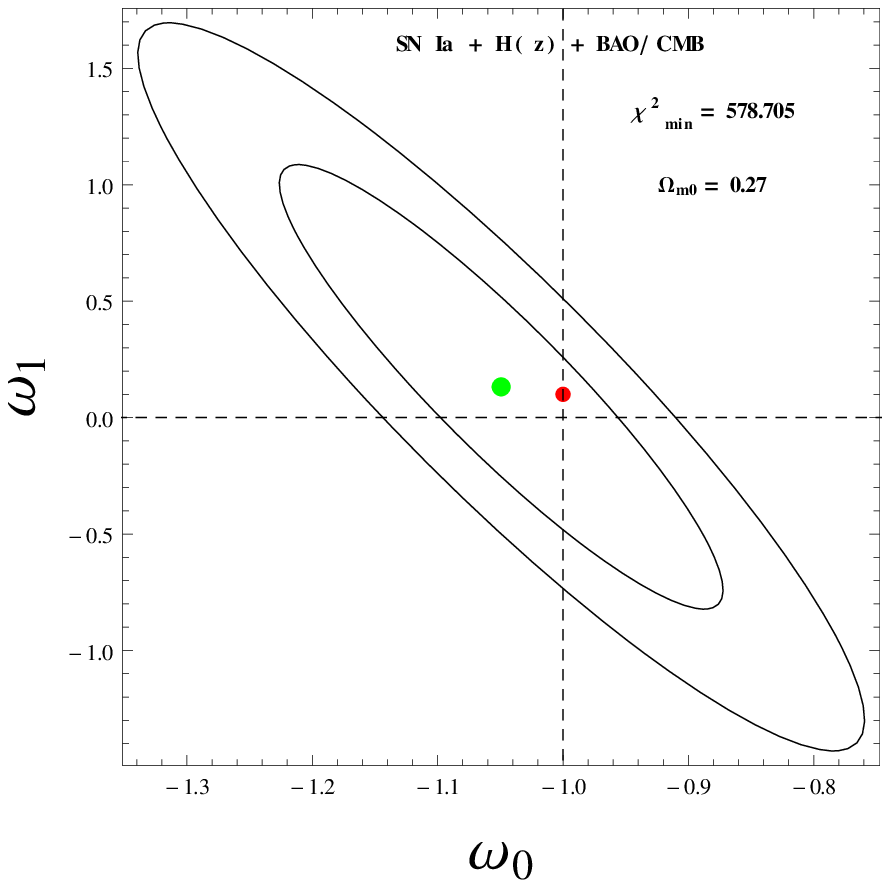,height=60mm,width=60mm}}
\centerline{\psfig{figure=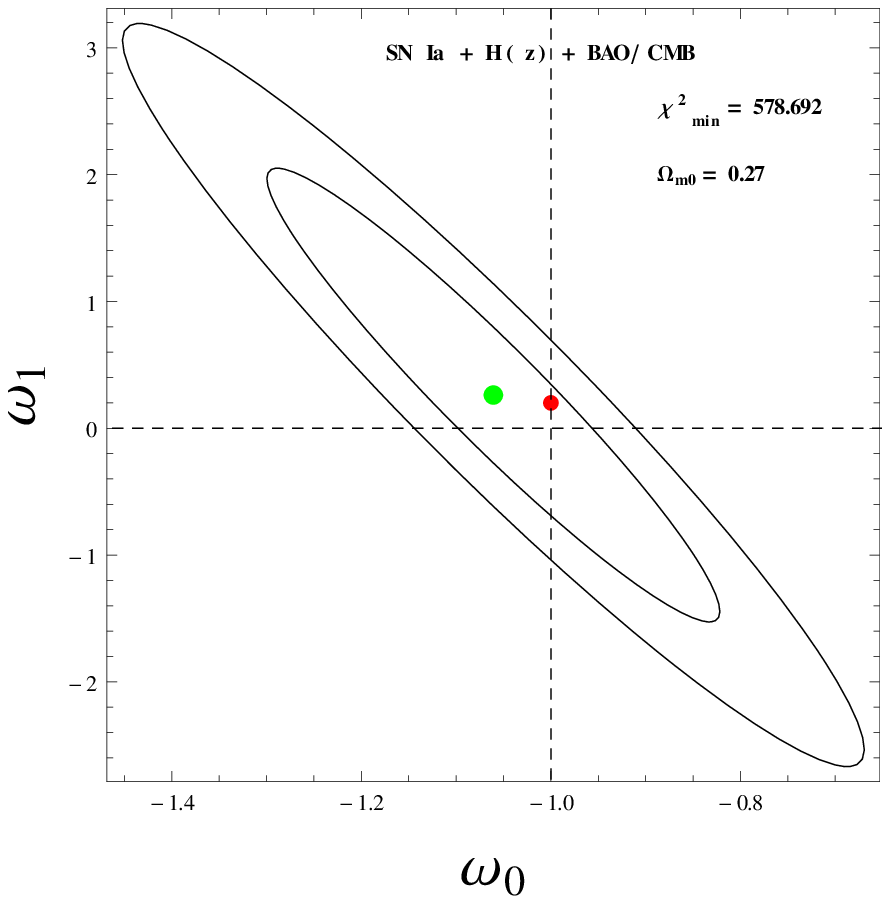,height=60mm,width=60mm}}
\centerline{\psfig{figure=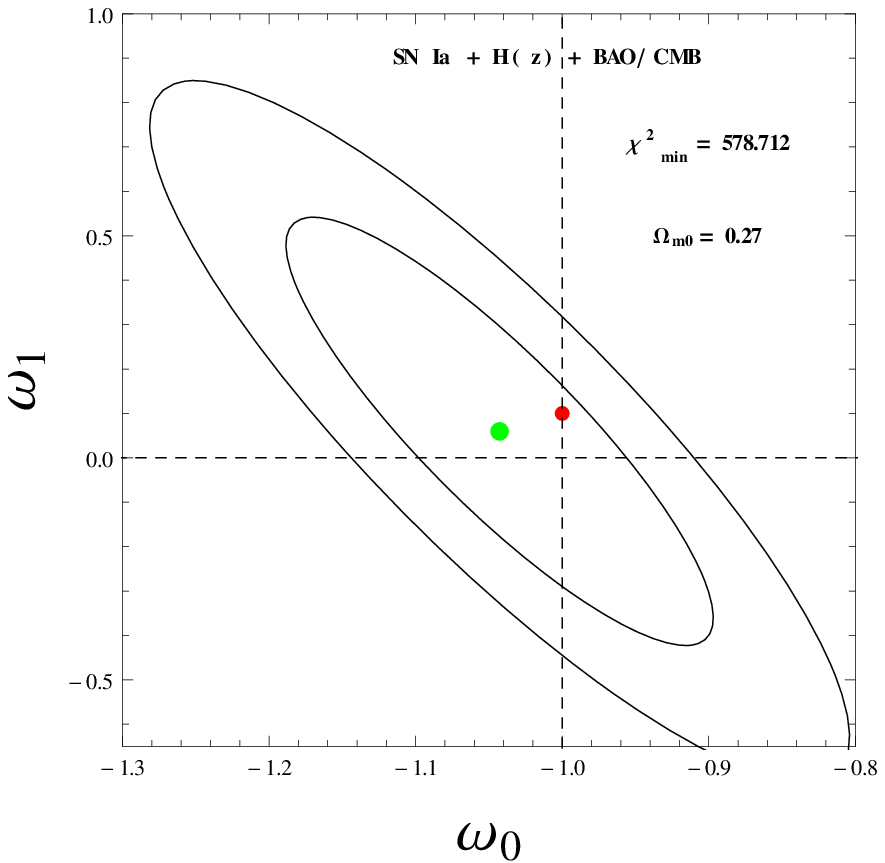,height=60mm,width=60mm}}
\caption{\normalsize{\em Plot of $1\sigma$ and $2\sigma$ confidence contours on  $\omega_{0} - \omega_{1}$ parameter space for the CPL parametrization (top panel), JBP parametrization (middle panel) and BA parametrization (bottom panel) respectively. In this graph, $\chi^2_{min}$ indicates the minimum value of $\chi^2$ corresponding to the best fit values of $\omega_{0}$ and $\omega_{1}$ for the SN Ia $+$ H(z) $+$ BAO/CMB dataset, as indicated in the frames. The fixed value of $\Omega_{m0}$ is also indicated in the frame.}}
\label{figcontour1}
\end{figure}
\begin{figure}[!h]
\centerline{\psfig{figure=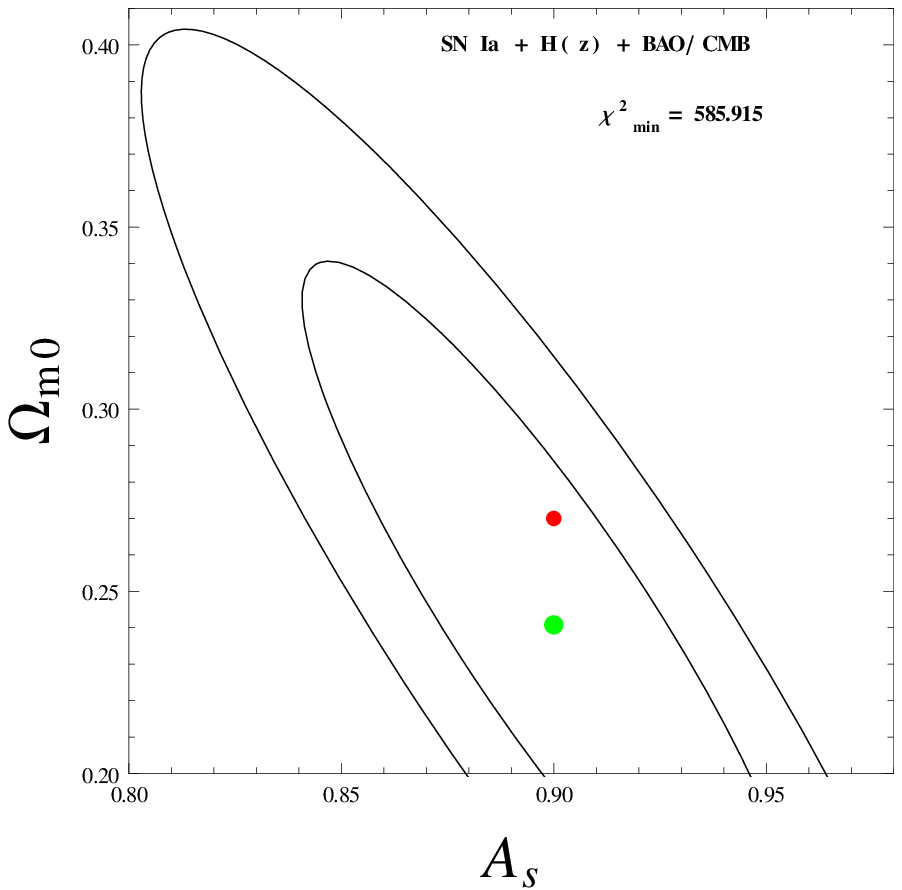,height=60mm,width=60mm}}
\caption{\normalsize{\em Plot of $1\sigma$ and $2\sigma$ confidence contours on  $A_{s} - \Omega_{m0}$ parameter space for the GCG parametrization. In this graph, $\chi^2_{min}$ indicates the minimum value of $\chi^2$ corresponding to the best fit values of $A_{s}$ and $\Omega_{m0}$ for the combined dataset (SN Ia $+$ H(z) $+$ BAO/CMB), as indicated in the frame. This is for $\alpha = -0.5$.}}
\label{figcontour2}
\end{figure}
\par In figure (\ref{figwphi}), we have shown the behavior of the EoS parameter $\omega_{\phi}(z) = \frac{\frac{1}{4}{\dot{\phi}}^4 - V(\phi)}{\frac{3}{4}{\dot{\phi}}^4 + V(\phi)}$, for four different parametrizations using the best fit values of the model parameters (as listed in Table \ref{table1} $\&$ \ref{table2}) for this dataset. It may be noted that the forms of EoS parameter (or potential) are different for different models. But, $\omega_{\phi}(z)$ $\approx$ $-1$ at present epoch ($z=0$ or $a=1$) for CPL, JBP $\&$ BA parametrization model, whereas for GCG model, $\omega_{\phi}(z)$ is around $-0.9$ at present but approach $-1$ value as evident from figure (\ref{figwphi}). Consequently, this feature suggests that ${\dot{\phi}}^4 << V(\phi)$ i.e., the potential term dominates over the non-canonical kinetic term to accelerate the cosmic expansion at present. It is also evident from figure (\ref{figwphi}) that the non-canonical scalar field behaves like phantom dark energy ($\omega_{\phi}<-1$, \cite{phantom1,phantom2,phantom3}) at present epoch for CPL, JBP and BA models. However, it deserves mention again that these plots of $\omega_{\phi}(z)$ are for the best-fit values of the model parameters and the results may change for the combination of other datasets. In this context, we would like to mention that the hypothetical phantom models meet several difficulties such as classical and quantum instabilities due to its negative kinetic energy and momentum as studied in refs. \cite{ppuzzles1, ppuzzles2, ppuzzles3, ppuzzles4}. In spite of that the recent observational data allow the possibility of the phantom EoS in the near past or in the near future \cite{jbp,phantomeos,trc, alam,alam01, feng}. So, from observational viewpoint, the phantom field one of the possible candidates for dark energy and can not be ruled out completely. Interestingly, it has been found that for the BA model, the EoS parameter  does not deviate much from $-1$ in future  (upto $z=-1$) and provides information regarding the complete evolution history of the universe but for CPL $\&$ JBP models however, the analysis is valid upto $z>-1$. The GCG behaves like  dark energy ($\omega_{\phi}=-0.9 > -1$) at late time and its equation of state also settles to a value close to $-1$ in the far future. So, like the GCG model, the BA model will also avoid the finite-time future singularity \cite{brip1,brip2,rip1,rip2,rip3,rip4}. It has also been noticed for the CPL and JBP models that the EoS parameter diverges in the finite future ($z \sim -1$) as expected from theoretical expressions of $\omega_{\phi}(z)$ for these models.\\
\par In the limit, $a (=\frac{1}{1+z})\rightarrow \infty$, we have $\rho_{\phi}(a)\rightarrow \infty$ and $p_{\phi}(a)\rightarrow \infty$ for CPL as well as JBP models. This implies that the universe (according to the CPL and JBP models) will end up in Big Rip singularity \cite{brip}, where the phantom energy density becomes very large in finite time and overcomes the gravitational repulsion. Despite this future divergency problem (at $z \sim -1$), the CPL $\&$ JBP parametrizations have several advantages such as, they can probe the past evolutionary history of the universe, have a bounded and well behaved behavior for high redshifts, projects a simple two parameter ($\omega_{0}$, $\omega_{1}$)-phase space with a good physical interpretation of $\omega_{0}$ and $\omega_{1}$, etc. In addition to the simplicity, this kind of parametrizations (CPL $\&$ JBP) also have the advantage of giving finite $\omega_{\phi}$ in the entire range $-1<z<\infty$. Furthermore, we would like to mention here that the two-parameter models, adopted in this paper, provides a simple choice to  obtain constraints from observational data.
\begin{figure}[!h]
\centerline{\psfig{figure=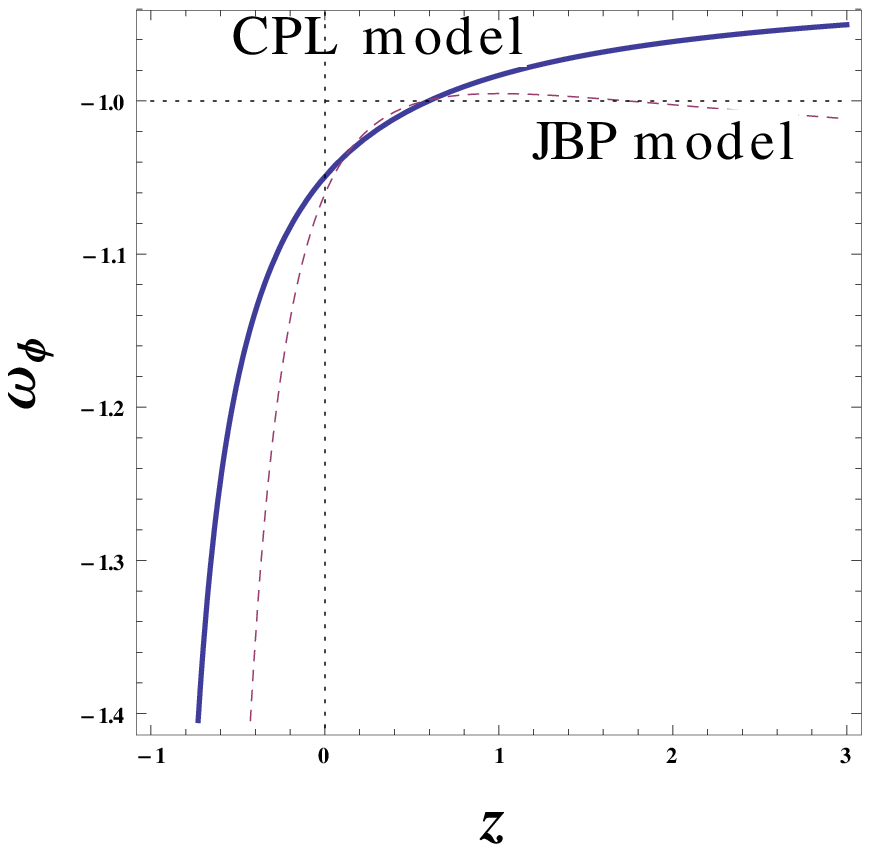,height=60mm,width=60mm}}
\centerline{\psfig{figure=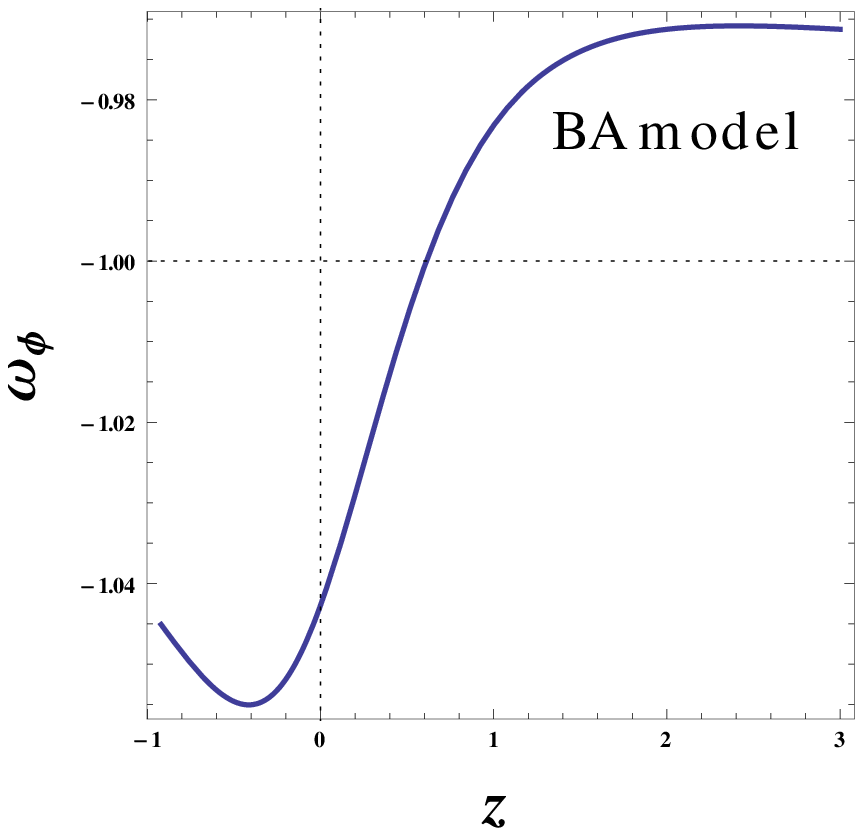,height=60mm,width=60mm}}
\centerline{\psfig{figure=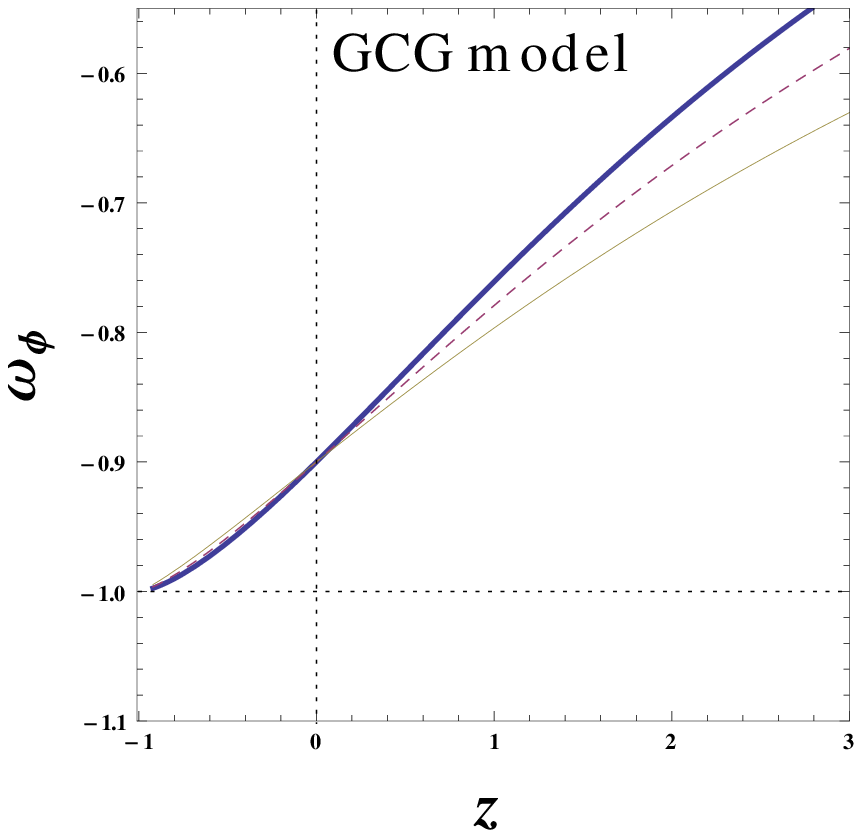,height=60mm,width=60mm}}
\caption{\normalsize{\em The top and middle panel represents the plot of the EoS parameter $\omega_{\phi}$ vs. redshift using the best fit values of ($\omega_{0}$, $\omega_{1}$) and $\Omega_{m0}=0.27$ (see Table \ref{table1}). The bottom panel corresponds to the evolution of $\omega_{\phi}$ for the GCG model. This plot is for the best fit values of $A_{s}$ and different values of $\alpha$ (see Table \ref{table2}); $\alpha=-0.5$ (thick curve), $\alpha=-0.55$ (dashed curve) and $\alpha=-0.6$ (thin curve). The intersection point of the dotted line indicates $\omega_{\phi}=-1$ at $z=0$, the $\Lambda$CDM case.}}
\label{figwphi}
\end{figure}
\section{Conclusion}
It is well known that the late-time accelerating expansion of the universe can be described by a scalar field. For this reason, in this present work, we have discussed about four different types of non-canonical scalar field models with varying dark energy EoS for understanding the observed cosmic expansion. As a time-dependent EoS plays an important role for understanding the nature of DE, so we have considered four phenomenological parametrizations of dark energy EoS. The dynamical features of each models are  analyzed, such as the evolutions of the deceleration parameter $q(z)$ and the density parameters ($\Omega_{\phi}$ and $\Omega_{m}$). The resulting cosmological behavior is found to be very interesting.\\
\par For all the toy models, it has been found that the deceleration parameter $q(z)$ indicates an early deceleration followed by a late time acceleration of the universe (see figures \ref{figq}a, \ref{figq1}a, \ref{figq2}a, \ref{figq3}a). We have also shown the evolution of density parameters and it is found that the results are in good agreement with recent observations \cite{omegaz,omegaz01}.\\
\par We have also compared our theoretical models with the observational data coming out of the latest SN Ia, Hubble parameter, BAO and CMB measurements. For this purpose, we have written the Hubble parameter $H(z)$ in terms of observable parameters ($z$, $H_{0}$ $\&$ $\Omega_{m0}$) and the corresponding model parameters for each DE parametrizations. We have obtained the best fit values of the parameters $\omega_{0}$ and $\omega_{1}$ by fixing the value of $\Omega_{m0}$ to $0.26$, $0.27$ and $0.28$ (shown in Table \ref{table1}). It may be important to mention here that the values of parameters of the model which were chosen for analytical results are well fitted in the $1\sigma$ and $2\sigma$ confidence contours for each parametrizations. We have found that the $\omega_{\phi}=-1$  crossing feature is also allowed by the SNIa $+$ H(z) $+$ BAO/CMB dataset for the CPL, JBP and BA models with its present best-fit EoS parameter, $\omega_{0}<-1$ (as presented in Table \ref{table1}). This is consistent with the results obtained by several authors \cite{jbp, phantomeos,trc, alam,alam01, feng}. On the other hand, we have also found that the standard $\Lambda$CDM model is still compatible at the $1\sigma$ confidence level for these toy models. It should be noted that the GCG model has been studied by many authors for the parameter range, $0\leq \alpha \leq 1$ \cite{cgas2,cgas3}. However, in this work, we have obtained the best fit values of the parameter $A_{s}$ and $\Omega_{m0}$ by fixing the value of other parameter $\alpha$ within the range, $-1 <\alpha < 0$ (shown in Table \ref{table2}), as $\alpha < 0$, is in good agreement with the work of Sen and Scherrer \cite{aas}, Gong et al. \cite{ygong} and Hazra et al. \cite{cpl7}. It has also been noticed from table \ref{table2} that the range of the allowed values of $\Omega_{m0}$ match well with the previous results obtained by Riess et al. \cite{omegaRange1} and Sahni et al. \cite{omegaRange2}.\\
\par However, as discussed in the previous section, the CPL and JBP models lose their prediction capability regarding the future evolution of the universe. We have also shown that it is very difficult to distinguish the GCG model from a $\Lambda$CDM in the near future and hence we need more investigations to constrain dark energy models more tightly. Obviously, we can not yet say which model is better as compared to other models by the analysis of the combined dataset (SN Ia $+$ H(z) + BAO/CMB). We hope that the next generation observational data can provide more tight constraints on EoS parameter to enrich our understanding regarding the nature of dark energy.
\section{Acknowledgements}
One of the authors (AAM) acknowledges UGC, Govt. of India for financial support through
Maulana Azad National Fellowship. SD wishes to thank IUCAA, Pune for warm hospitality where part of this work has been carried out. The authors are also thankful to the anonymous referee whose valuable comments have helped in improving the quality of this paper.

\end{document}